\documentclass[twoside,12pt]{article}
\usepackage{amssymb,amsmath}
\usepackage[dvipdfmx]{graphicx}
\usepackage{color}
\usepackage[breaklinks, colorlinks=true]{hyperref}

\newcommand{\be}{\begin{equation}}
\newcommand{\ee}{\end{equation}}
\newcommand{\bea}{\begin{eqnarray}}
\newcommand{\eea}{\end{eqnarray}}

\newcommand{\Tc}{T_{\text{c}}}
\newcommand{\rme}{\mathrm{e}}
\newcommand{\rmi}{\mathrm{i}}
\newcommand{\rmd}{\mathrm{d}}

\newcommand{\tr}{\mathrm{tr}}
\newcommand{\bA}{\boldsymbol{A}}
\newcommand{\bB}{\boldsymbol{B}}
\newcommand{\bD}{\boldsymbol{D}}
\newcommand{\bE}{\boldsymbol{E}}
\newcommand{\bH}{\boldsymbol{H}}
\newcommand{\ez}{\hat{\boldsymbol{e}}_z}

\newcommand{\bj}{\boldsymbol{j}}

\newcommand{\bp}{\boldsymbol{p}}

\newcommand{\bx}{\boldsymbol{x}}

\newcommand{\boldu}{\boldsymbol{u}}
\newcommand{\bS}{\boldsymbol{S}}
\newcommand{\bM}{\boldsymbol{M}}
\newcommand{\bL}{\boldsymbol{L}}
\newcommand{\bJ}{\boldsymbol{J}}
\newcommand{\bmu}{\boldsymbol{\mu}}
\newcommand{\bnabla}{\boldsymbol{\nabla}}
\newcommand{\bgamma}{\boldsymbol{\gamma}}
\newcommand{\bsigma}{\boldsymbol{\sigma}}
\newcommand{\bSigma}{\boldsymbol{\Sigma}}
\newcommand{\bomega}{\boldsymbol{\omega}}
\newcommand{\btau}{\boldsymbol{\tau}}
\newcommand{\calA}{\mathcal{A}}
\newcommand{\calB}{\mathcal{B}}
\newcommand{\calC}{\mathcal{C}}
\newcommand{\calD}{\mathcal{D}}

\newcommand{\calK}{\mathcal{K}}
\newcommand{\calL}{\mathcal{L}}

\newcommand{\calP}{\mathcal{P}}

\newcommand{\calT}{\mathcal{T}}

\newcommand{\tiln}{\tilde{n}}
\newcommand{\tilp}{\tilde{p}}

\newcommand{\feyn}[1]{
  \setbox0=\hbox{\ensuremath{#1}}
  \hbox to\wd0{\hbox to0pt{\hbox to\wd0{\hss/\hss}\hss}\box0}}
\renewcommand{\Re}{\mathrm{Re}}
\renewcommand{\Im}{\mathrm{Im}}
\newcommand{\comment}[1]{}  

\begin{document}
\title{Extreme matter in electromagnetic fields and rotation}
\author{Kenji Fukushima\\
\\
{\small Department of Physics, The University of Tokyo,
  7-3-1 Hongo, Bunkyo-ku, Tokyo 113-0033, Japan}}
\maketitle

\begin{abstract}
  We look over recent developments on our understanding about
  relativistic matter under external electromagnetic fields and
  mechanical rotation.  I review various calculational approaches for
  concrete physics problems, putting my special emphasis on generality
  of the method and the consequence, rather than going into
  phenomenological applications in a specific field of physics.  The
  topics covered in this article include static problems with magnetic
  fields, dynamical problems with electromagnetic fields, and
  phenomena induced by rotation.
\end{abstract}
\tableofcontents

\section{Introduction}
\label{sec:intro}

This review is not intended to be a comprehensive chapter of
encyclopedia but a collection of selected topics on relativistic
matter (i.e., matter with fermions having relativistic energy
dispersion relations) affected by external electromagnetic fields
and/or rotation, which should have a wide variety of applications in
various physics contexts.  In this review I would not go into
phenomenological applications but will put my emphasis on
methodologies, assuming that readers are already engaged in some
physics problems with electromagnetic fields and/or rotation, and are
rather interested in practical approaches.  If one is interested in
applications in condensed matter physics, one is invited to read
Refs.~\cite{RevModPhys.73.629,Miransky:2015ava}.  In the context of
the high-energy nuclear reactions, Ref.~\cite{Hattori:2016emy}
provides us with the state-of-the-art phenomenological progresses
among which, particularly, the topologically induced effects are
nicely summarized in an earlier review of
Ref.~\cite{Kharzeev:2015znc}.

Although I put aside phenomenological discussions, I would briefly
mention where to find physical targets, so that general readers can be
more cognizant of backgrounds and motivations.  In condensed matter
physics the magnetic field has been the best probe to topological
aspects, and examples include the (fractional) quantum Hall effect and
the chiral magnetic effect (CME).  In fact, some class of quantum Hall
plateaus under strong magnetic fields may be ascribed to the magnetic
catalysis, i.e., an abnormal enhancement of condensation in the scalar
channel.  Interestingly, the ideas of the magnetic catalysis as well
as the CME were born originally in the context of high-energy physics,
and then, they found applications in condensed matter setup.
Actually, in the present universe, the strongest magnetic field is a
human-made one;  in laboratory experiments of positively charged
nucleus-nucleus collisions at high energy, simulations estimate as
strong magnetic fields as $\sim 10^{14}\,$T or even stronger, which is
thousands stronger than the surface magnetic field of the magneter, a
special kind of the neutron star with gigantic magnetic fields.  Even
though the impulse strength is such huge, the life time of the
magnetic field is extremely short of order of $\sim 10^{-23}\,$s, and
interpretations of the experimental data are still under debates.
Since the collision dynamics is furiously changing in time, not only
magnetic fields but also electric fields are as strong, leading to
production of particles from the vacuum.  Such a phenomenon of
insulation breakdown of the vacuum is long known as the Schwinger
Mechanism and one of the modern challenges in laser physics is to
reach the Schwinger limit.  The Schwinger Mechanism is also essential
to understand microscopic descriptions of the aforementioned chiral
magnetic effect.  Unlike the magnetic field that decays quickly in the
nuclear reactions, people in high-energy physics realized that the
rotation may stay longer because of the angular momentum conservation,
and they start thinking of the vorticity as an alternative (and more
promising) probe to topological phenomena.  Possible interplays
between the rotation, the magnetic field, and the finite fermionic
density lead to new topological implications, and interestingly, the
neutron stars possess all features of the rotation, the magnetic
field, and the finite density.  Accumulation of neutron star
observation data is expected to give us some clues for anomalous
manifestations in astrophysical condensed matter systems in the
future.  Now, I hope, readers see that the topics chosen in the
present review are not randomly picked up, but are tightly connected
to each other.

This introduction, in the present review, will be followed by three
sections focused on three different but related topics.  In the next
section we will explicitly see instructive and convenient calculus to
deal with external magnetic field $\bB$.  The most remarkable and
universal feature of relativistic matter is the magnetic catalysis
induced by constant $\bB$, on which this present review cannot avoid
having some overlap with other articles, see, e.g.,
Ref.~\cite{Miransky:2015ava}.  In this review we will elucidate, on
top of direct calculation, the renormalization group argument to
deepen our intuitive understanding on the magnetic catalysis.

Furthermore, in addition to the well-known (inverse) magnetic
catalysis, we will address two important but technically involved
topics to upgrade oversimplified setups to more realistic situations.
One is the effect of inhomogeneous $\bB(\bx)$ and the other is the
effect of boundary imposed for finite size systems.  These are
distinct effects definitely, but they have similarity to some extent,
leading to modifications on the Landau levels.

For inhomogeneity, fortunately, a special profile of the magnetic
field is known, for which the Dirac equation can be analytically
solved.  This special case is often referred to as the Sauter-type
potential problem.  We will see a similar Sauter-type problem when we
discuss electric fields later.  Thanks to this solvable example, we
can acquire some useful insights about how spatial inhomogeneity
should modify the conventional pattern of the Landau degeneracy.  Such
rigorous results are valid only for a specific example of $\bB(\bx)$,
but we may well expect to learn general qualitative modifications
triggered by inhomogeneous $\bB(\bx)$.  The Landau degeneracy would be
lifted up also by a different physical disturbance, that is, the
finite size effect even for homogeneous magnetic fields.  We will show
explicit calculations with a boundary in the cylindrical coordinates.
We will go into rather technical details here;  some of the
expressions will be useful in later discussions on rotation.

Here, in this review, we would not consider time-dependent $\bB(t)$.
For static magnetic fields we can utilize time-independent vector
potentials, and the ground state of matter under such static magnetic
fields may stably exist.  It is a theoretically intriguing problem to
identify the ground state structures as functions of static magnetic
strength.  Then, naturally, one might think that a time-dependent
$\bB(t)$ should be far more challenging and interesting enough to
deserve closer discussions.  However, there are (at least) two reasons
for our sticking to static magnetic fields at present.  The first and
crucial reason is that a nonperturbative analysis with time-dependent
$\bB(t)$ is impossibly difficult, while a diagrammatic method may work
perturbatively if $\bB$ is sufficiently weak or $\bB(t)$ behaves like
a sharp pulse to justify the impulse approximation.  There is no
Sauter-type solvable problem known for time-dependent $\bB(t)$.  One
might wonder if one can anyway solve the Dirac equation numerically
instead of pursuing an analytical solution, but such an even
brute-force calculation would suffer principle difficulty.  Numerical
calculations inevitably rely on lattice discretization and a certain
scheme of boundary condition.  The possible magnetic flux is quantized
in order to satisfy imposed boundary condition.  Therefore,
continuously changing $\bB(t)$ as a function of time would be
impossible in such lattice discretized systems.  The second reason for
our focusing on time-independent $\bB$ in this review is that we would
like to address the effect of electric field $\bE$ mainly from the
dynamical point of view.  The difference between $\bB$ and $\bE$
appears from quantum numbers of these fields; $\bB$ is $\calT$-odd and
$\calP$-even, and $\bE$ is $\calT$-even and $\calP$-odd.  This means
that the time-derivative, $\dot{\bB}$, is $\calT$-even as is $\bE$,
resulting in similar dynamical evolutions, though they are opposite in
the parity.  So, if we place an electrically charged particle into a
system under either $\dot{\bB}$ or $\bE$, the field gives a finite energy to
the particles and accelerates them.  More interestingly, if the given
energy exceeds a mass threshold, an onshell pair of a particle and an
anti-particle is created, which is called the Schwinger Mechanism.

The next section after static magnetic field is devoted to discussions
on the Schwinger Mechanism for spatially homogeneous electric field.
For a comprehensive review on the Schwinger Mechanisms and the
effective Lagrangians, see Ref.~\cite{Dunne:2004nc}.  If
electromagnetic fields themselves are time-dependent, the pair
production of a particle and an anti-particle is simply an inverse
process of the annihilation.  What is particularly remarkable about the
Schwinger Mechanism is that even a static $\bE$ is intrinsically
dynamical, while a static $\bB$ cannot transfer any work onto charged
particles.  To concentrate on effects solely induced by $\bE$, as we
mentioned above, we will limit ourselves to the case with
time-independent $\bB$ throughout this review.

There are many formulations and derivations for the particle
production rate associated with the Schwinger Mechanism.  In this
review I will present somehow explicit calculations using the
Sauter-type potential for time-dependent $\bE(t)$, which also includes
static $\bE$ as a special
limit.  In principle, the strategy to derive the pair production rate
can be generalized to arbitrary electromagnetic backgrounds with help
of some numerical calculations.  Although there are tremendous
progresses in the numerical techniques (for such an example, see
Ref.~\cite{Gelis:2015kya}), we will concentrate only on the analytical
approaches in this review.  For this purpose I will give detailed
explanations on a saddle-point approximation within the framework of
the worldline formalism.  The worldline formalism is based on
Schwinger's proper-time integration.  Interestingly, in the
saddle-point approximation, the sum over multi-particle production
corresponds to the sum over the winding number which classifies
classical solutions of the equation of motion.  Thus, the
calculational machinery employed with the saddle-point approximation
is commonly referred to as the ``worldline instanton'' approximation.

The presence of both $\bE$ and $\bB$ with $\bE\cdot\bB\neq 0$
provides us with the simplest ``optical'' setup to probe a non-trivial
sector with respect to the chiral anomaly (where I use a word,
optical, to mean, not visual light, but externally controllable
electromagnetic waves in general).  When the fermion mass $m$
is vanishing, the axial vector current is expected to be conserved,
but quantum fluctuations give rise to a violation of the would-be
current conservation.  That is, the divergence of the axial vector
current is no longer zero but is proportional to $\bE\cdot\bB$ if
there are those background fields.  We shall pay our attention to the
chiral anomaly using the proper-time integration, to notice that
taking the $m\to 0$ limit turns out to be a subtle procedure.  With
concrete calculations we will demonstrate the importance to
distinguish the in-state and the out-state once the system involves
background $\bE$;  the expectation values of physical observables are
calculable with the in-in expectation values, while the in-out
expectation values are amplitudes whose squared quantities correspond
to the expectation values of physical observables.  As an application
of the in- and out-state calculus we will cover an optical realization
of the chiral anomaly.  One clear signature for the chiral anomaly is
an anomalous contribution to the electric resistance called the
negative magnetoresistance.  The microscopic picture is established as
a manifestation of the chiral magnetic effect (CME) (which is one of
central subjects of Ref.~\cite{Kharzeev:2015znc};  see also
Ref.~\cite{Fukushima:2012vr} for a memorial essay on the CME).

Finally, we shall turn to discussions on effects induced by mechanical
rotation.  There are two possibly equivalent but technically very
different approaches to deal with rotation;  one is the
field-theoretical calculation in a rotating frame, for which the
rotation is global, and the other is the fluid dynamical description
with vorticity vectors, for which the rotation is local.  If the
local vorticity vector uniformly distributes over two-dimensional
space, it eventually amounts to the rigid rotation, as is the case for
the vortex lattice of a rotating superfluid (see, for example,
Ref.~\cite{pierre1999superconductivity}).  Therefore, these two
treatments should be in principle equivalent, and nevertheless, each
method has some advantages and disadvantages.  The advantage to use the
local vorticity vector is that only the local properties are concerned
and it is not necessary to impose a boundary not to break the
causality.  However, the price to pay is, one must know a correct
theoretical reduction to fluid dynamics in such a way to keep
essential features of the chiral anomaly.  Such a reduction program
is, to some extent, successful, leading to the chiral kinetic theory
and the anomalous hydrodynamics.

Solving the Dirac equation in a rotating frame is a rather brute-force
method, but the advantage is that one does not have to worry about a
theoretical reduction to slow variables.  Once the solutions of the
Dirac equation are given, all necessary information must be contained
in those solutions.  In this review I take my preference to proceed
into technical details on the field-theoretical treatments in a
rotating frame.  In this way we will see a derivation of the chiral
vortical effect (CVE), which is an analogue of the CME with the
magnetic field replaced by the rotation (see
Ref.~\cite{Kharzeev:2015znc} for more details).  Such an explicit
calculation will give us a useful insight into the microscopic origin
of the CVE -- unlike the quantum anomaly which typically originates
from the ultraviolet edges of the momentum integration, the CVE
emerges from a finite discrepancy between a continuous integral and a
discrete sum, which is rather analogous to the Casimir effect on the
technical level.

Another interesting feature of rotating system is found in the
application of the Floquet theory which states a mathematical theorem
on differential equations with periodic dependence (see, e.g.,
Refs.~\cite{Hanggi:1997,doi:10.1080/00018732.2015.1055918} for
pedagogical reviews).  We can regard electromagnetic fields in a
rotating frame as time periodically driven forces, and several
techniques are known to tackle such Floquet-type problems.  It is
impossible to cover the whole stories about the Floquet theory, and so
we will take a quick view of a technology called the Magnus expansion,
which is a common method to make the Floquet problem well-defined from
the high-frequency limit.  Also, we point out that the Floquet theory
has a fascinating interpretation as a higher (spatial) dimensional
theory with an external electric field whose strength is related to
the frequency.  I will explain this interesting correspondence from a
motivation that our knowledge on dynamical problems with
electromagnetic fields could have an interdisciplinary application to
consider the Floquet-type problem.  In other words, instead of
imposing an external electric field, one may think of shaking a system
time periodically to mimic the effect of electric field.  This idea
might remind us of the synthetic magnetic field realized by the
Floquet engineering (see Ref.~\cite{RevModPhys.83.1523} for a review)
but the interpretation of electric field is rather simpler.

The last topics discussed in this review are relativistic extensions
of the gyromagnetic effects, namely, the Barnett effect and the
Einstein--de~Haas effect (for an established and the most comprehensive
review, see Ref.~\cite{RevModPhys.7.129}).  It is a widely known
notion that the mechanical rotation and the magnetization can be
converted to each other;  precisely speaking, the orbital angular
momentum can be converted to the spin due to the spin-orbit coupling,
and vice versa.  This conversion itself is a quite robust process,
which implies that the relativistic extensions be straightforward.  In
relativistic theories, however, the separation of the total angular
momentum into the orbital and the spin components is ambiguous.  It
is thus still under disputes whether the relativistic Barnett and
Einstein--de~Haas effects can exist as straightforward generalization
of classical descriptions.  I will introduce some interesting
theoretical speculations, hoping that this part of the present review
will be followed up by future research.

Before closing the introduction, let us make sure conventions assumed
for the present review.  Throughout this review, we consistently use
words, ``static'', ``homogeneous'', and ``constant'' in the following
way.  Static $\bE$ and $\bB$ mean time-independent fields, i.e.,
$\dot{\bE}=0$ and $\dot{\bB}=0$.  Homogeneous fields are independence
of spatial coordinates, that is, $\partial_x \bE=0$ and $\partial_x
\bB=0$ where $x$ is any of spatial coordinates.  For static and
homogeneous electromagnetic fields, we often call them constant.  In
Minkowskian spacetime our choice of coordinates is $(x^0,x^1,x^2,x^3)$
and $x_0=x^0$, $x_1=-x^1$, $x_2=-x^2$, $x_3=-x^3$.  We often use $x$,
$y$, $z$ to denote $x^1$, $x^2$, $x^3$.  For momenta we employ similar
conventions and $p_x$, $p_y$, $p_z$ often indicate $p^1$, $p^2$,
$p^3$ (not $p_1$, $p_2$, $p_3$).  This rule is applied also for $\bE$ and $\bB$.  In Euclidean
spacetime ($x^4,x^1,x^2,x^3$) and ($x_4,x_1,x_2,x_3$) are just
indistinguishable.

Finally, we note that $\hbar=c=1$ and $\varepsilon_0=\mu_0=1$ in the
natural unit.  Moreover, in this review, we will not treat the
in-medium electromagnetic fields, so we will thoroughly employ the
electric field $\bE$ and the magnetic flux $\bB$, but never use the
electric flux $\bD$ nor the magnetic strength $\bH$ in this review.
(There is no difference for the vacuum fields in the natural unit.)

\section{Static Problems with Magnetic Fields}
\label{sec:magnetic}

The Maxwell equations have duality between the electric field $\bE$
and the magnetic field $\bB$, which is often referred to as
Heaviside-Larmor symmetry.  Since the temporal and the spatial
directions are not equivalent in Minkowskian spacetime, however, there
is a tremendous difference in effects of $\bE$ and $\bB$ upon matter.
In most of our considerations in the present review we limit ourselves
to static $\bE$ and $\bB$.  One might then think of static problems of
matter under such static $\bE$ and $\bB$;  however, a static $\bE$
would induce an electric current (if matter is not an insulator), and
strictly speaking, such a system with a sustained current can never be
equilibrated but it is a steady state.  In contrast, a static $\bB$
would not transfer any work on charged particles, and one can consider
an equilibrated system even at finite but static $\bB$, which is our
main subject of this section.

\subsection{Calculus with constant magnetic field}
\label{sec:calculusB}

It is a textbook knowledge how to solve the Schr\"{o}dinger equation
in constant $\bB$ to derive the Landau quantization.  There is,
however, almost no textbook that explains how to solve the Dirac
equation with constant $\bB$.  Some complication appears from the fact
that the Dirac spinors have components with different spin
polarizations which respond to $\bB$ differently.  In this subsection
we make a quick summary of the most naive calculations of solving the
Dirac equation in detail.  As we will see in the next subsection, for
practical purposes, an alternative formulation based on the
proper-time integration turns out much more convenient.  Nevertheless,
explicit calculations are quite instructive, as explicated in this
subsection.

\subsubsection{Solutions of the Dirac equation}
\label{sec:direct}

We here review a direct method to solve the Dirac equation as a
straightforward extension from the standard technique for the
Schr\"{o}dinger equation.  In the literature this method is often
called the Ritus method~\cite{RITUS1972555} (see also
Refs.~\cite{Noronha:2007wg,Fukushima:2009ft} for Ritus projections in
concrete problems).  We can fix the direction of $\bB$ along the
$z$-axis without loss of generality, and shall choose a gauge,
$A^0=A^1=A^3=0$ and $A^2=Bx$, corresponding to
$\bB=\bnabla\times \bA=B\ez$.  In the following we always assume
$eB>0$ to avoid putting modulus and simplify the expressions.  Then,
let us introduce two functions using the harmonic oscillator
wave-functions:
\begin{equation}
  \begin{split}
    f_{n,p}^+(x) &\equiv \rme^{-\rmi(\omega t - p_y y - p_z z)}
    \phi_n(x - p_y/eB) \qquad (n=0,1,\dots) \\
    f_{n,p}^-(x) &\equiv \rme^{-\rmi(\omega t - p_y y - p_z z)}
    \phi_{n-1}(x - p_y/eB) \qquad (n=1,2,\dots)
  \end{split}
\end{equation}
and $f_{0,p}^-(x)=0$.  For the harmonic oscillator wave-functions the
frequency is characterized by the magnetic scale as
\begin{equation}
  \phi_n(x) \equiv \sqrt{\frac{1}{2^n n!}} \biggl(\frac{eB}{\pi}\biggr)^{1/4}
  e^{-\frac{1}{2} eB x^2} H_n(\sqrt{eB}\, x)\,,
\end{equation}
where $H_n(x)$ represents the Hermite polynomials of degree $n$.  Now,
let us introduce a $4\times 4$ projection matrix as
\begin{equation}
  P_{n,p}(x) \equiv \frac{1}{2} \bigl[ f_{n,p}^+(x) + f_{n,p}^-(x) \bigr]
  + \frac{i}{2} \bigl[ f_{n,p}^+(x) - f_{n,p}^-(x) \bigr]\gamma^1 \gamma^2\,,
\end{equation}
which is reduced to
$P_{n=0,p}(x)=P_+$ for the lowest Landau level with
$P_\pm\equiv\frac{1}{2}(1\pm\rmi\gamma^1\gamma^2)$.
After several line calculations one can easily find the following
relation,
\begin{equation}
  \bigl( \rmi\feyn{\partial} - e\feyn{A} - m \bigr) P_{n,p}(x)
  = P_{n,p}(x) \bigl( \omega\gamma^0 + \sqrt{2eBn}\,\gamma^2
  - p_z \gamma^3 - m \bigr) \,.
\end{equation}
At this point, using the above relation, we can readily write down the
solutions of the Dirac equation.  Changing the order of the Dirac
operator and the projection operator makes the right-hand side be a
form of the free Dirac equation with the momenta replaced as
$p\to \tilp=(\omega,0,-\sqrt{2eBn},p_z)$.  Thus, the solutions of the Dirac
equation at finite constant $\bB$ read,
\begin{equation}
  \psi(n,p;x) = P_{n,p}(x)\, u(\tilp,s)\,,\qquad
  \psi(n,p;x) = P_{n,p}(x)^\ast\, v(\tilp,s)
\end{equation}
for particles and anti-particles.  Here, the explicit expressions for
spinors in momentum space are,
\begin{equation}
  u(\tilp,s) = \frac{1}{\sqrt{\omega+m}} \begin{pmatrix}
      (\omega + m)\xi_s \\ \bsigma\cdot\tilde{\bp}\, \xi_s
    \end{pmatrix}\,,\qquad
  v(\tilp,s) = \frac{1}{\sqrt{\omega+m}} \begin{pmatrix}
    \bsigma\cdot\tilde{\bp}\, \eta_s \\ (\omega+m)\eta_s
    \end{pmatrix}
  \label{eq:spinors}
\end{equation}
in the Dirac representation of the $\gamma$ matrices.  As usual,
$\xi_s$ and $\eta_s$ are two-component spin bases satisfying
$\xi_s^\dag \xi_s=\eta_s^\dag \eta_s = 1$ (with no sum over $s$), and
$\bsigma$'s are Pauli matrices.  For perturbative calculations the
free propagator is the most elementary building block of Feynman
diagrams, which can be immediately constructed with the solutions of
the Dirac equation.  The propagator then has such an explicit form of
the momentum integration and the Landau level sum as
\begin{equation}
  S_0^F(x,y) = \langle T \psi(x)\bar{\psi}(y) \rangle
  = \int\frac{\rmd \omega\,\rmd p_y\,\rmd p_z}{(2\pi)^3} \sum_n
  P_{n,p}(x) \frac{\rmi\,(\feyn{\tilp} + m)}
  {p_\parallel^2 -2eB n - m^2 + \rmi\epsilon} P_{n,p}^\ast(y)\,.
\label{eq:prop_direct}
\end{equation}
Here, $p_\parallel\equiv (\omega,0,0,p_z)$.  It is important to make
several remarks here about the above propagator.  Sometimes I see a
bit misleading statement about $\tilp$ as if $\tilp$ were a genuine
momentum flowing on a fermionic propagator, but such a naive picture
would violate the momentum conservation at vertices of the Yukawa
coupling.  Since the integrand in the above expression contains,
$\rme^{\rmi p_y(x^2-y^2)}$, from $P_{n,p}(x)$ and $P_{n,p}^\ast(y)$,
it is not $\tilde{p}$ but $p_y$ which enters the momentum
conservation, even though it does not show up in the energy dispersion
relation.  I would make a next comment about the translational
invariance.  For constant $\bB$ the system should keep the
translational invariance, which implies that $S_0^F(x,y)$ be a 
function of $x-y$ alone, but the vector potential $\bA$ is $x$
dependent and it seemingly violates the translational invariance.  In
fact, obviously, $S_0^F(x,y)$ is not a function of $x-y$, but as we
will confirm soon later, it is only a phase that breaks the
translational invariance.

We note that the propagator in the lowest Landau level approximation
(LLLA) obtains from $n=0$, that is,
\begin{equation}
  \begin{split}
  S_{\rm LLLA}^F(x,y) &= \int \frac{\rmd \omega\,\rmd p_y\,\rmd p_z}
  {(2\pi)^3}\; \rme^{-\rmi \omega(x^0-y^0)+\rmi p_y(x^2-y^2)
    +\rmi p_z(x^3-y^3)} \\
  &\qquad\qquad\times \sqrt{\frac{eB}{\pi}}\;
  \rme^{-\frac{1}{2}eB[(x^1-p_y/eB)^2+(y^1-p_y/eB)^2]}
  \frac{\rmi\,(\feyn{p}_\parallel + m) P_+}{p_\parallel^2 - m^2 + \rmi\epsilon}\,.
  \end{split}
\end{equation}
From this form it is obvious again that $S_{\rm LLLA}^F(x,y)$ is not a
function of $x-y$.  Interestingly, we can easily separate the
translational invariant and the non-invariant parts after the $p_y$
integration, leading to the following expression,
\begin{equation}
  \begin{split}
  S_{\rm LLLA}^F(x,y) &= \rme^{\rmi\frac{eB}{2}(x^1+y^1)(x^2-y^2)} \cdot
  \frac{eB}{2\pi}\,
  \rme^{-\frac{eB}{4}[(x^1-y^1)^2 + (x^2-y^2)^2]} \\
  &\qquad\qquad\times
  \int \frac{\rmd \omega\,\rmd p_z}
  {(2\pi)^2}\; \rme^{-\rmi \omega(x^0-y^0)+\rmi p_z(x^3-y^3)}
  \frac{\rmi\,(\feyn{p}_\parallel + m) P_+}
       {p_\parallel^2 - m^2 + \rmi\epsilon}\,.
  \end{split}
\end{equation}
This overall phase factor is nothing but the Aharonov-Bohm (AB) phase
by $\exp(-\rmi e\int_y^x \rmd z_\mu A^\mu)$ with the vector potential
representing $\bB$.  For most loop calculations the phase factors
cancel out, and it is convenient to introduce the Fourier
transformation of the phase removed part,
$\tilde{S}(x-y)\equiv \rme^{\rmi e\int_y^x \rmd z_\mu A^\mu} S(x,y)$.  The
translational invariant part of the LLLA propagator in momentum space
thus reads,
\begin{equation}
  \tilde{S}_{\rm LLLA}^F(p) = \frac{\rmi\,(\feyn{p}_\parallel + m)
  P_+\, 2\rme^{-p_\perp^2/eB}}
        {p_\parallel^2 - m^2 + \rmi\epsilon}\,,
\label{eq:prop_LLLA}
\end{equation}
where $p_\perp^2\equiv p_x^2+p_y^2$.  The last exponential damping
factor is important;  the well-known Landau degeneracy factor,
$eB/(2\pi)$, appears from the integrations over $p_x$ and $p_y$ which
are convergent thanks to this exponential damping factor.

Indeed, it is important to note that the transverse momentum dependence
appears only through $\rme^{-p_\perp^2/eB}$, so that the integration over
transverse phase space is separately performed, leading to the Landau
degeneracy factor,
\begin{equation}
  \int\frac{\rmd^2 p_\perp}{(2\pi)^2}\,2\rme^{-p_\perp^2/eB}
  = \frac{eB}{2\pi}\,.
\end{equation}
Now that the transverse integration is done, the (3+1)-dimensional
dynamics is subject to restricted space along the $t$- and $z$-axes,
that is, the system is effectively reduced to (1+1) dimensions.  We
must emphasize that the particle motion in configuration space is not
restricted at all, but there is no energy cost with transverse
motions, which results in the dimensional reduction.  A short summary
on the LLLA is:
\[
\framebox{
  \begin{minipage}{0.8\textwidth}
    LLLA: An approximation under strong magnetic fields to reduce the
    (3+1)-dimensional phase space of the Dirac fermions into the
    transverse Landau degeneracy factor, $eB/(2\pi)$, and
    (1+1)-dimensional dynamics along the magnetic direction.
  \end{minipage}
}
\]
This approximation is justified as follows;  as seen from the
denominator in Eq.~\eqref{eq:prop_direct} the Dirac fermions are
gapped by $2eBn$.  If $eB$ is larger than typical energy scales of the
sytem, higher excited states with $n\neq 0$ are far apart from the
lowest Landau level, and the LLLA works quite well.  In any case,
whether it is a good approximation or not, the LLLA is always useful
for us to understand the magnetic effect intuitively.

\subsubsection{Schwinger proper-time method}
\label{sec:proper-time}

Next, let us see how to find the same expression from a completely
different passage which is much shorter than solving the Dirac
equation directly.

Schwinger developed a useful method~\cite{Schwinger:1951nm} for
general electromagnetic backgrounds.  The idea is that, if we are
interested in the propagator only, we do not have to solve the Dirac
equation but what we should do is just to take an inverse of the Dirac
operator with the electromagnetic fields contained in the covariant
derivative.  This is sufficient for most of practical purposes;  as we
will discuss later, a scalar-operator condensate is a physical
quantity of our interest to demonstrate the magnetic catalysis.  Any
such operators are to be expressed as a combination of the
propagators.

Now, an important step for the actual calculation in Schwinger's
method is that the operator inverse can be expressed in an integral
form with inclusion of an auxiliary variable, $s$, which is called the
proper time.  That is, with a notation of $D_\mu$ for the covariant
derivative, we can write,
\begin{equation}
  S_0^F(x,y) = \langle x| \frac{\rmi(\rmi\feyn{D}+m)}
  {-\feyn{D}^2 - m^2 + \rmi\epsilon} |y\rangle
  = (\rmi\feyn{D}_x + m) \int_0^\infty \rmd s\,
  \langle x| \exp[\rmi (-\feyn{D}^2-m^2+\rmi\epsilon)s] |y\rangle\,.
\label{eq:S0proper}
\end{equation}
We note that $\feyn{D}^2$ produces a matrix,
$\exp(-\frac{\rmi}{2}eF\cdot\sigma s)$ with
$F\cdot\sigma = F_{\mu\nu}\sigma^{\mu\nu}$, where the spin tensor is
defined as usual;
$\sigma^{\mu\nu}\equiv \frac{\rmi}{2}[\gamma^\mu,\gamma^\nu]$.  Then,
it is an easy exercise to find,
\begin{equation}
  \rme^{-\frac{\rmi}{2}eF\cdot \sigma s} = 
  \cos(eB s) \bigl[ 1 + \gamma^1\gamma^2 \tan(eB s) \bigr]\,.
\end{equation}
In this way we can continue sorting out the expression, and after
all, we can extract $\tilde{S}_0^F(x-y)$, which can be Fourier
transformed into $\tilde{S}_0^F(p)$ given
by~\cite{Schwinger:1951nm,Gusynin:1994xp,Gusynin:1995nb},
\begin{equation}
  \begin{split}
  \tilde{S}_0^F(p) &= \int_0^\infty \rmd s\,
  \exp\biggl[ -\rmi \biggl(m^2 + \omega^2 + p_z^2 + p_\perp^2
    \frac{\tan(eB s)}{eB s} \biggr)s \biggr] \\
  &\qquad \times \bigl[ \feyn{p}+m + (p_y\gamma^1-p_x\gamma^2)
    \tan(eB s) \bigr] \cdot \bigl[ 1 + \gamma^1\gamma^2
    \tan(eB s) \bigr]\,.
  \end{split}
\label{eq:Stilde}
\end{equation}
This expression for the propagator is so useful that we will later
revisit this including the electric field when we discuss the
worldline instanton approximation and also the axial Ward identity.
It is not easy, however, to implement the LLLA from the above
expression.  Interestingly, it is possible to reexpress the above
propagator into the following
form~\cite{Gusynin:1995nb,Fukushima:2012kc},
\begin{equation}
  \tilde{S}^F(p) = \sum_{n=0}^\infty \frac{\rmi\,\{ (
    \feyn{p}_\parallel+m)\bigl[ P_+ A_n(p_\perp^2)
      + P_- A_{n-1}(p_\perp^2) \bigr] + \feyn{p}_\perp B_n(p_\perp^2)
    \}}{p_\parallel^2 - 2eBn - m^2 + \rmi\epsilon}\,,
\label{eq:prop_full}
\end{equation}
where the sum over $n$ corresponds to the Landau level sum.  The
numerator consists of two functions,
$A_n(p_\perp^2) \equiv 2\rme^{-2z}(-1)^n L_n^{(0)}(4z)$ and 
$B_n(p_\perp^2) \equiv 4\rme^{-2z}(-1)^n L_{n-1}^{(1)}(4z)$ with
$z\equiv p_\perp^2/(2eB)$.  Here, $L_n^{(\alpha)}(x)$ denotes the
generalized Laguerre polynomials defined by
\begin{equation}
  L_n^{(\alpha)}(x) \equiv \frac{\rme^x x^{-\alpha} }{n!}
  \frac{\rmd^n}{\rmd x^n}\, \bigl(\rme^{-x} x^{n+\alpha}\bigr)\,.
\end{equation}
From this expanded form, we can deduce the LLLA propagator immediately
from $n=0$, which exactly reproduces Eq.~\eqref{eq:prop_LLLA}.
Usually, if $eB$ is large enough as compared to other energy scales in
the problem, the LLLA works excellently to capture the essence, as we
already stated before.

\subsection{Magnetic Catalysis}
\label{sec:catalysis}

Now we are ready to understand what the magnetic catalysis is, which
is a phenomenon of abnormal enhancement of a scalar condensate.  The
magnetic catalysis has an impact of paramount importance in nuclear
physics.  The origin of particle masses is known to be spontaneous
breakdown of chiral symmetry, as revealed first by Nambu and
Jona-Lasinio~\cite{Nambu:1961tp}, and it is nothing but a scalar
condensate that causes symmetry breaking.   This implies that a scalar
condensate catalyzed by imposed magnetic field makes particle masses
increased, which is a conclusion from the magnetic catalysis.

Let us turn to technicalities now.  We can write the scalar condensate
in terms of the propagator as
\begin{equation}
  \Sigma = \langle \bar{\psi}\psi \rangle
  = -\lim_{y\to x}\tr S(x,y) = -\lim_{z\to 0}\tr\tilde{S}(z)\,.
\label{eq:scalar}
\end{equation}
This quantity measures a condensate formed with a particle and an
anti-particle (or a hole at finite density) and is generally induced
if the interaction is strong enough.  From the physics point of view,
$\bar{\psi}\psi$ is an operator conjugate to the mass, since the mass
term in the Dirac Lagrangian density is $m\bar{\psi}\psi$, which means
that $\Sigma\neq 0$ would play a role as a source term to induce a
larger effective mass.  In this section we will quantify the relation
between the bare masses and the effective masses in the presence of
the scalar condensate.

\begin{figure}
  \centering
  \includegraphics[width=0.2\textwidth]{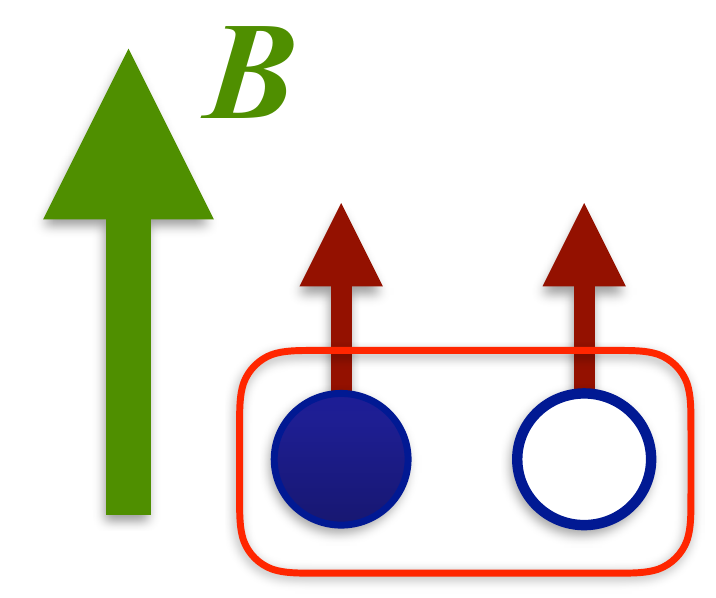}
  \caption{Schematic illustration for magnetic enhancement of the
    scalar condensate.}
  \label{fig:scalar}
\end{figure}

Condensates or the vacuum expectation values of physical operators should
generally depend on external parameters such as the temperature, the
density, and the electromagnetic background fields.  It is thus an
interesting question how $\Sigma$ changes when a finite $\bB$ is
turned on as studied in a pioneering work of
Ref.~\cite{Suganuma:1990nn}.  Intuitively, it is naturally anticipated
that $\bB$ would favor a formation of the scalar condensate from the
following argument.  To assign proper quantum numbers to a composite
state of a particle and an anti-particle, in the non-relativistic
language, the scalar condensate must have the orbital angular momentum
$L=1$ to cancel the intrinsic parity which is opposite for a particle
and an anti-particle.  Then, $S=1$ is required to make the total
angular momentum $J=0$.  Even though the whole quantity,
$\bar{\psi}\psi$, is charge neutral and has no direct coupling to
$\bB$, from the microscopic level, we may well expect that such a
spin-triplet configuration could be intrinsically more favored by
$\bB$, which is schematically illustrated in Fig.~\ref{fig:scalar}.

In most interested cases, $\Sigma(B)$, i.e., the scalar condensate as
a function of magnetic strength $B=|\bB|$ should increase with
increasing $B$.  Such increasing behavior is actually the case in the
strong interaction physics in which the low-energy theorem holds
including the magnetic effect~\cite{Shushpanov:1997sf}, which was
revisited more extensively in Ref.~\cite{Cohen:2007bt}.  Although this
tendency itself is rather robust not sensitive to specific theory, the
precise dependence on $B$ should be different for different
interactions and degrees of freedom contained in the theory.  In the
literature, this general tendency of increasing $\Sigma$ with
increasing magnetic strength $B$ is commonly referred to as the
magnetic catalysis.

In the original context~\cite{Gusynin:1995nb,Gusynin:1994re}, however,
the magnetic catalysis is defined in a more strict manner.  It has
been found that the magnetic field plays a role as a catalyst to cause
nonzero $\Sigma$ for even infinitesimal interaction coupling.  The
statement shall be:
\[
\framebox{
  \begin{minipage}{0.8\textwidth}
    Magnetic Catalysis: The scalar condensate takes a nonzero
    expectation value for even infinitesimal attractive coupling if
    the system is placed under sufficiently strong magnetic field.
  \end{minipage}
}
\]
Roughly speaking, it is caused by low-dimensional nature;  the system
under strong magnetic fields is effectively reduced to (1+1)
dimensions.  Then, the gap equation for $\Sigma$ exhibits the same
structure as that for the gap energy in superconductivity for which
the Fermi surface realizes one dimensionality.  In fact, plugging
Eq.~\eqref{eq:Stilde} into Eq.~\eqref{eq:scalar}, performing the
four-momentum integration which yields $eB s^{-1}$ to compensate for
the mass dimension 4, we finally get,
\begin{equation}
  \Sigma \simeq -\frac{eB}{4\pi^2} m \int_{1/\Lambda^2}^\infty \rmd s\,
  s^{-1}\,\rme^{-\rmi m^2 s} \cot(eB s)
  \simeq -\frac{eB}{4\pi^2} m \Gamma[0,m^2/\Lambda^2]\,,
\end{equation}
where a ultraviolet cutoff $\Lambda$ is adopted and $\Gamma[a,x]$
denotes the incomplete gamma function.  To arrive at the last
expression, the integration contour is rotated as $\rmi s\to s$ and
then $\coth(eB s)$ is approximated to be unity for large $eB$, which
corresponds to the LLLA{}.  Because
$\Gamma[0,x]\sim -\gamma_{\rm E}-\ln x$ for small $x$, we understand
that $\Sigma$ should behave like $\sim m\ln m$ for small $m$.  The
operator $\bar{\psi}\psi$ is conjugate to $m$, so that $\Sigma$ is
obtained by the $m$-derivative of the energy $E[m]$.  Therefore, such
behavior of $\Sigma\sim m\ln m$ means that $E[m]$ should contain a
term $\sim m^2\ln m$.  This observation will be soon verified in the
subsection where we will see a model calculation of $\Sigma$ as a
function of $B$ and the coupling constant $\lambda$.

\subsubsection{Direct calculation}

My goal in this subsection is to present direct calculations employing
an interacting model.  The model is defined with the following
Lagrangian density:
\begin{equation}
  \calL = \bar{\psi}\rmi\feyn{\partial}\psi
  + \frac{\lambda_\Lambda}{2} \bigl[ (\bar{\psi}\psi)^2
    + (\bar{\psi}\rmi\gamma_5\btau\psi)^2 \bigr]\,,
\end{equation}
for which the bare mass is assumed to be vanishing, $m=0$, and
$\lambda_\Lambda$ is the coupling at the ultraviolet scale $\Lambda$.
This type of model is commonly called the Nambu--Jona-Lasinio (NJL)
model~\cite{Nambu:1961tp} (for extensive reviews on the NJL model, see
Refs.~\cite{Klevansky:1992qe,Hatsuda:1994pi}), which is a relativistic
extension of the BCS model.

In the mean-field approximation, a quantum field is decomposed as
$\bar{\psi}\psi = (\bar{\psi}\psi - \langle\bar{\psi}\psi\rangle)
+\langle\bar{\psi}\psi\rangle$, with which higher-order fluctuations
in $\bar{\psi}\psi-\langle\bar{\psi}\psi\rangle$ are neglected.  In
this approximation the mean-field Lagrangian density reads:
\begin{equation}
  \calL_{\rm MF} = \bar{\psi}\rmi\feyn{\partial}\psi
  +\lambda_\Lambda \langle\bar{\psi}\psi\rangle
  -\frac{\lambda_\Lambda}{2}\langle\bar{\psi}\psi\rangle^2
  = \bar{\psi}(\rmi\feyn{\partial}-M)\psi - \frac{M^2}{2\lambda_\Lambda}\,.
\end{equation}
Now we understand that $\langle\bar{\psi}\psi\rangle$ gives rise to
an effective mass term, namely,
$M=-\lambda_\Lambda\langle\bar{\psi}\psi\rangle$.  To determine the
energetically favored value of $M$, we need to evaluate the energy $E$
as a function of $M$, that is given in the mean-field approximation by
\begin{equation}
  E[M] = -2\int_{p^2 \le \Lambda^2}\frac{\rmd^3 p}{(2\pi)^3}
  \sqrt{p^2 + M^2} + \frac{M^2}{2\lambda_\Lambda}\,,
\label{eq:NJL_E}
\end{equation}
where $\Lambda$ is the ultraviolet cutoff scale.  The first term
represents the zero-point oscillation energy, and the second is the
condensation energy from the mean field.  Introducing a dimensionless
variable, $\xi\equiv M/\Lambda$, the energy can be expanded as
\begin{equation}
  E[\xi] \simeq -\frac{\Lambda^4}{4\pi^2}(1+\xi^2)
  + \frac{\Lambda^2}{2\lambda_\Lambda}\xi^2 + O(\xi^4)\,.
\end{equation}
The trivial vacuum at $\xi=0$ becomes unstable when the quadratic
coefficient in front of $\xi^2$ is negative.  There are two competing
effects for this coefficient;  the zero-point oscillation energy
always has a negative coefficient favoring a nonzero $\xi$, while the
condensation contribution has the opposite sign.  Therefore, if the
coupling $\lambda_\Lambda$ is large enough, the condensation energy is
suppressed, and the zero-point oscillation energy overcomes to lead to
a finite $\xi$ or $M$.  Thus, the spontaneous generation of dynamical
mass $M\neq 0$ requires a large enough $\lambda_\Lambda$, i.e.,
\begin{equation}
  \lambda_\Lambda \Lambda^2 > 2\pi^2\,.
  \label{eq:SSB_cond}
\end{equation}
We need higher-order terms $\sim O(\xi^4)$ to locate an optimal value
of $\xi$.  The above calculations are relatively simple, and yet,
surprisingly, this successfully accounts for the origin of masses.

Let us repeat the same procedure for the system with strong magnetic
field.  For simplicity the magnetic field is assumed to be strong
enough to justify the LLLA in which the phase space integration is
replaced by the Landau degeneracy factor and the (1+1)-dimensional
integration:
\begin{equation}
  2\int_{p^2\le\Lambda^2} \frac{\rmd^3 p}{(2\pi)^3} \;\;
  \longrightarrow \;\;
  \frac{eB}{2\pi} \int_{p^2\le\Lambda^2} \frac{\rmd p_z}{2\pi}\,.
\end{equation}
We note that there is only one spin degrees of freedom in the LLLA as
a result of the projection by $P_+$, and so the overall spin factor 2
does not appear in the right-hand side in the LLLA above.  Then, we
can compute the energy as
\begin{equation}
  E[M] = -\frac{eB}{2\pi}\int_{-\Lambda}^\Lambda
  \frac{\rmd p_z}{2\pi}\,\sqrt{p_z^2+M^2} +
  \frac{M^2}{2\lambda_\Lambda}
  \simeq -\frac{eB\Lambda^2}{4\pi^2}\biggl[ 1+\Bigl(
    \ln\frac{2}{\xi} + \frac{1}{2} \Bigr)\xi^2 \biggr]
  +\frac{\Lambda^2}{2\lambda_\Lambda}\xi^2 + O(\xi^4)\,.
\end{equation}
This form of the expanded energy is extremely interesting.  As we
promised, a term $\sim M^2 \ln M$ appears.  Thus, the quadratic
coefficient is $M$ or $\xi$ dependent, and for $\xi\ll 1$, the
coefficient involving $\ln(2/\xi)$ becomes arbitrarily large, which
means that the negative coefficient can always overcome to make
$\xi\to 0$ unstable for even infinitesimal coupling
$\lambda_\Lambda$.  Therefore, some $\xi\neq 0$ is energetically
favored as long as $eB$ and $\lambda_\Lambda$ are non-vanishing, and
the condition~\eqref{eq:SSB_cond} in this case is changed to
$\lambda_\Lambda > 0$.  From the above expression of $E[M]$,
interestingly, we can find an energy minimum even without $O(\xi^4)$
term.  Some calculational steps eventually reach,
\begin{equation}
  M_0 = \xi_0\Lambda = 2\Lambda\,\exp\biggl(
  -\frac{2\pi^2}{\lambda_\Lambda\, eB} \biggr)\,.
\end{equation}
The analogy to superconductivity is evident in view of the above
expression.  The $\lambda_\Lambda$ dependence enters the gap as an
essential singularity around $\lambda_\Lambda = 0$, which is common to
the BCS gap energy.  In this way, at the same time, we see that
particle masses are not physical constants but they exhibit ``medium''
dependence influenced by external magnetic fields.

\subsubsection{Renormalization group analysis}

It would be quite instructive to consider an alternative approach to
understand the magnetic catalysis based on the renormalization group
(RG) equation following discussions in
Refs.~\cite{Scherer:2012nn,Fukushima:2012xw}.  This RG criterion is a
very powerful technique, and even when it is difficult to solve the
gap equation in the broken phase, we can make an educated
guess about the parametric dependence of the gap.  This method is
generally known as the Thouless criterion.

\begin{figure}
  \centering
  \includegraphics[width=0.8\textwidth]{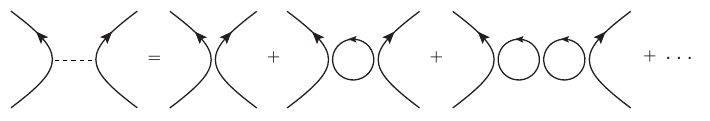}
  \caption{RG resummation diagrams for the four fermionic vertex.}
  \label{fig:RG_diag}
\end{figure}

The point is that the four fermionic coupling $\lambda_k$ is running
with scale $k$ in the RG language;  $\lambda_k$ is the coupling after
modes over $k\sim\Lambda$ are integrated out.  The bare coupling in
the Lagrangian density takes the ultraviolet value, $\lambda_\Lambda$
at the scale $\Lambda$, and $\lambda_k$ should receive loop
corrections sketched in Fig.~\ref{fig:RG_diag} as $k$ goes smaller.

From a one-loop diagram we can find the $\beta$ function for
$\lambda_k$ as
\begin{equation}
  k \partial_k \lambda_k = -\frac{\lambda_k^2 k^2}{3\pi^2}\,.
\label{eq:beta}
\end{equation}
We can directly solve this differential equation with the initial
condition of $\lambda_\Lambda$ at $k=\Lambda$, leading to
\begin{equation}
  \lambda_k = \frac{\lambda_\Lambda}
         {1+\dfrac{\lambda_\Lambda}{6\pi^2}(k^2 - \Lambda^2)}\,,
\end{equation}
which increases as $k$ goes smaller.  If the above is expanded in
terms of $\lambda_\Lambda$, the geometric series indeed correspond to loop
corrections in Fig.~\ref{fig:RG_diag}.  In the infrared limit, $k=0$,
the denominator is $1-\lambda_\Lambda \Lambda^2/(6\pi^2)$.  Hence,
$\lambda_k$ diverges at some point of $k$ before reaching $k=0$ if
\begin{equation}
  \lambda_\Lambda \Lambda^2 > 6\pi^2\,.
\label{eq:SSB_cond_RG}
\end{equation}
The divergent coupling implies that the scattering amplitude of a
particle and an anti-particle has a singularity which signifies a
formation of the bound state, like a formation of the Cooper pair in
superconductivity.  The above inequality is thus interpreted as the
condition~\eqref{eq:SSB_cond} in the mean-field approximation.  The
discrepancy by a factor 3 supposedly comes from different cutoff
schemes.

\begin{figure}
  \centering
  \includegraphics[width=0.4\textwidth]{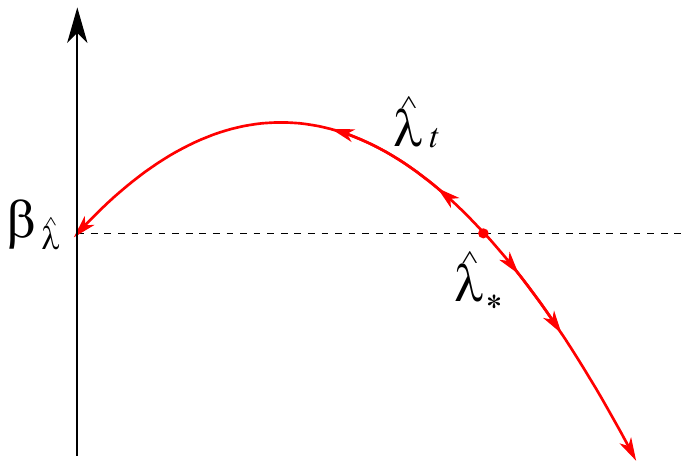} \hspace{1em}
  \includegraphics[width=0.4\textwidth]{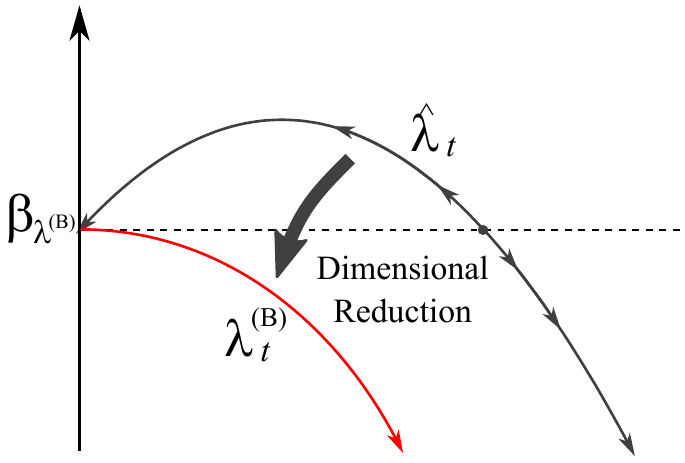}
  \caption{RG running of the coupling constant without magnetic field
    (left panel) and under strong magnetic field (right panel).
    Figures taken from Ref.~\cite{Fukushima:2012xw}.}
  \label{fig:run}
\end{figure}

There is a more appealing way to understand the critical value of the
coupling strength using the dimensionless coupling,
$\hat{\lambda}_k\equiv\lambda_k k^2$.  This rescaling to make $\lambda_k$
dimensionless corresponds to the rescaling procedure in the
conventional RG (Kadanoff) transformation.  Then, the $\beta$ function
for the rescaled coupling is translated into
\begin{equation}
  \beta_{\hat{\lambda}} = k\partial_k \hat{\lambda}_k = 2\hat{\lambda}_k
  - \frac{\hat{\lambda}_k^2}{3\pi^2}\,,
\label{eq:dless_beta}
\end{equation}
where the first term is added from the mass dimensionality of the
coupling $\lambda_k$.  Then, the zero of the $\beta$ function is
located at $\hat{\lambda}_\ast = 6\pi^2$.  If the RG running is
initiated from $\hat{\lambda}_\Lambda < \hat{\lambda}_\ast$, then
$\beta_{\hat{\lambda}}$ is positive, and so $\hat{\lambda}_k$ becomes
smaller as $k$ gets smaller, which is schematically illustrated in the
left panel of Fig.~\ref{fig:run}.  On the other hand, if the running
is launched from $\hat{\lambda}_\Lambda > \hat{\lambda}_\ast$, then
$\beta_{\hat{\lambda}}$ is negative and $\hat{\lambda}_k$ keeps
increasing with decreasing $k$ and it eventually diverges as shown in
the left panel of Fig.~\ref{fig:run}.  In this way, from graphical
analysis, we can identify the critical coupling in
Eq.~\eqref{eq:SSB_cond_RG} from the zero of $\beta_{\hat{\lambda}}$
without solving the differential equation.

Now that we established our understanding without magnetic field, let
us apply this argument to matter under strong magnetic field.  We
adopt the LLLA and replace the transverse phase space integration by
the Landau degeneracy factor, which modifies Eq.~\eqref{eq:beta} into
\begin{equation}
  \partial_t \lambda_k = - \frac{eB}{2\pi^2} \lambda_k^2\,,
\end{equation}
which can be directly solved as
\begin{equation}
  \lambda_k = \frac{\lambda_\Lambda}
         {1+\dfrac{\lambda_\Lambda\,eB}{2\pi^2}\, t}\,.
\end{equation}
Here, $t\equiv\ln(k/\Lambda)$ ranges from $0$ (at $k=\Lambda$) to
$-\infty$ (at $k=0$), and so the denominator can become zero for
sufficiently large negative $t$ if only $\lambda_\Lambda eB>0$.  This
is precisely what is expected from the magnetic catalysis.  As we did
for $\hat{\lambda}_k$, we can define dimensionless coupling,
$\lambda^{\rm (B)}_k \equiv (eB/2\pi^2)\lambda_k$.  In this case
associated $\beta_{\lambda^{\rm (B)}}$ is as simple as
\begin{equation}
  \beta_{\lambda^{\rm (B)}} = \partial_t\lambda^{\rm (B)}_t
  = -(\lambda^{\rm (B)}_t)^2\,,
\label{eq:dless_beta2}
\end{equation}
for which the zero is found only at $\lambda^{\rm (B)}_\Lambda=0$.
This drastic difference from Eq.~\eqref{eq:dless_beta} is
attributed to the transverse phase space.  For
Eq.~\eqref{eq:dless_beta}, by definition, the dimensionless coupling
contains $k^2$ running with $k$, but for Eq.~\eqref{eq:dless_beta2}
the magnetic field is a more relevant scale.  Then, $k^2$ running is
lost due to the dimensional reduction from (3+1)- to (1+1)-dimensional
dynamics.  We see that $\beta_{\lambda^{\rm (B)}}$ in
Eq.~\eqref{eq:dless_beta2} is entirely negative, which means that
$\lambda^{\rm (B)}_{t\to-\infty}$ inevitably diverges for any initial
$\lambda_\Lambda$.  That is, the magnetic catalysis is again concluded
from the dimensional reduction.

\subsection{Inverse Magnetic Catalysis}
\label{sec:inverse}

The magnetic field has a general tendency to favor the scalar
condensate, as we have discussed, but sometimes the resulting effect
appears opposite.  Such an exceptional situation, i.e., decreasing
behavior of the scalar condensate for increasing magnetic field, is
called the inverse magnetic catalysis, and the first example was found
in a system at finite density~\cite{Preis:2010cq}.  If the density is
high enough, in fact, it is always the case that the scalar condensate
does not increase but decreases with increasing magnetic
field~\cite{Preis:2012fh}, which is universally confirmed also in the
RG analysis~\cite{Fukushima:2012xw}.

Nowadays, the inverse magnetic catalysis refers to a different
realization, namely, a situation of finite temperature environment in
quantum chromodynamics~\cite{Bali:2011qj}, which is a non-Abelian
Yang-Mills theory with several flavors of fermions.  This version of
the inverse magnetic catalysis at finite temperature depends on
microscopic dynamics of theory and there is no simple way to explain
underlying mechanism.  Nevertheless, since the recognition of the
inverse magnetic catalysis posed lots of theoretical problems in the
field of high-energy nuclear physics, we will go through to make some
remarks on the finite-temperature phase transitions.

In short, the inverse magnetic catalysis can be defined as
\[
\framebox{\begin{minipage}{0.8\textwidth}
    Inverse Magnetic Catalysis: The scalar condensate is decreased
    if the external magnetic field is applied to the system and its
    strength is increased.  This may happen due to interplay with
    other external parameters and/or other dynamics of the theory
    affected by the magnetic field.
  \end{minipage}
}
\]

\subsubsection{Finite density}

The scalar condensate is suppressed at finite density or finite
chemical potential $\mu$.  We shall consider the effect of finite
density using the NJL model again.  At finite density the energy
function, Eq.~\eqref{eq:NJL_E}, is slightly modified as
\begin{equation}
  E[M;\mu] = -\int_{p^2\le\Lambda^2}\frac{\rmd^3 p}{(2\pi)^3}\,
  \biggl( \bigl|\sqrt{p^2+M^2}-\mu\bigr|
    + \bigl|\sqrt{p^2+M^2}+\mu\bigr| \biggr)
  + \frac{M^2}{2\lambda_\Lambda}\,.
\end{equation}
One may notice that, if $M$ is larger than $\mu$, there is no $\mu$
dependence at all, and this should be of course so.  As long as $\mu$
is small not exceeding the mass threshold, no finite-density
excitation is allowed, and thus no density effect should be visible.
In what follows below we assume $\mu > M$, and then we can relax the
modulus in the above energy expression, simplifying it into
\begin{equation}
  E[M;\mu] =
  -2\int_{p^2\le p_{\rm F}^2}\frac{\rmd^3 p}{(2\pi)^3}\,
  \bigl(\mu - \sqrt{p^2+M^2} \bigr)
  -2\int_{p^2\le\Lambda^2}\frac{\rmd^3 p}{(2\pi)^3}\,
  \sqrt{p^2+M^2}
  + \frac{M^2}{2\lambda_\Lambda}\,,
\end{equation}
where $\mu = \sqrt{p_{\rm F}^2+M^2}$.  We can easily make sure that
the finite-$\mu$ correction by the first term produces a term,
$p_{\rm F}^2 M^2/(12\pi^2)$, at the quadratic order in $M$, which
energetically favors $M=0$.  When a magnetic field is imposed, the
density of states is changed;  continuous excited levels become
shrunk into the discrete Landau levels with degeneracy.  Therefore,
this finite-$\mu$ term to favor $M=0$ is enhanced at larger magnetic
field since the density of states is squeezed by the magnetic effect.
In summary, in this case of finite density, the inverse magnetic
catalysis occurs as a result of finite-$\mu$ effect enhanced by the
magnetic field.  There is no theoretical difficulty;  the inverse
magnetic catalysis is to be observed already in the mean-field
approximation.

\subsubsection{Finite temperature}

The NJL model is a relativistic cousin of the BCS theory.  In this
type of the theory the critical temperature, $\Tc$, is proportional to
the gap, $\Sigma$, at zero temperature, i.e., $\Tc\propto\Sigma(T=0)$,
where $\Tc$ is defined by the condition, $\Sigma(\Tc)=0$.  Because the
magnetic catalysis with larger $B$ drives
$\Sigma=\langle\bar{\psi}\psi\rangle$ to a larger value, finite-$T$
calculations using the NJL model predict that the melting temperature,
$\Tc$, where $\Sigma=0$ is reached, monotonically increases with
increasing $B$.

In the context of the strong interaction with quarks and gluons, such
a possibility to shift $\Tc$ of the system exposed to strong $B$ has
attracted a lot of theoretical interest.  Quarks and gluons interact
nonperturbatively, and it is believed that the color charge is
confined and the effective mass is dynamically generated in the low
temperature phase.  At high temperature the coupling constant runs
with the energy scale and the system enters a weak-coupling regime,
where both color confinement and dynamical mass generation are lost.
Because these phenomena are clearly distinct, there is no necessity
for two temperatures of deconfinement and melting to coincide.
Numerical simulations in the first-principles theory of the strong
interaction have revealed, however, that two phenomena occur almost
simultaneously at one critical temperature.

This observation of the common critical temperature strongly suggests
that the origins of confinement and dynamical mass generation are not
distinct but should be traced back to some common microscopic
mechanism in the first-principles theory.  As an attempt to model a
locking mechanism underlying confinement and mass generation, the NJL
model has been augmented to what is called the PNJL
model~\cite{Fukushima:2003fw,Ratti:2005jh} including not only $\Sigma$
but also another order parameter corresponding to confinement.  The
PNJL model in a strong magnetic field predicted disentanglement of two
phenomena; $\Sigma$ is affected by the magnetic catalysis, while
confinement belongs to the dynamics of gluons which are electric
charge neutral, and thus the deconfinement temperature is hardly
changed by the magnetic effect.  If this is the case, the phase
diagram of strongly interacting matter out of quarks and gluons would
open a new window in which confinement is lost but fermions are still
massive~\cite{Mizher:2010zb,Gatto:2012sp}.

\begin{figure}
  \centering
  \includegraphics[width=0.4\textwidth]{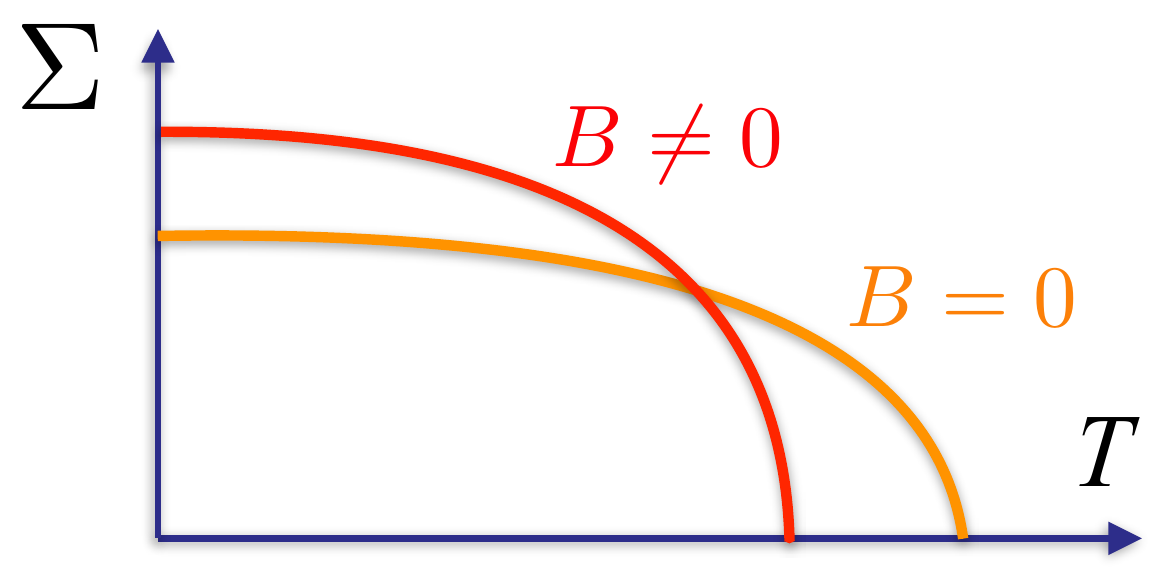}
  \caption{Schematic illustration of the change in $\Sigma(T)$ caused
    by the magnetic field.  The orange color line is the original
    behavior at $B=0$, which is pushed up to a larger value at small
    $T$.  As $T$ grows larger, however, $\Sigma(T)$ drops steeper and
    hits zero at smaller critical $T$ as shown by the red color line.}
  \label{fig:inverse_mc}
\end{figure}

With this background stories in mind, one may imagine how much
surprised people were at numerical results from the first-principles
simulation claiming that there is no disentanglement but the common
phase transition temperature is lowered by a stronger magnetic
field~\cite{Bali:2011qj,Bali:2012zg,Bali:2013esa}.
This means, if the order parameter $\Sigma$ is plotted as a function
of $T$ for zero and nonzero magnetic field, $\Sigma$ gets larger at
small $T$, but $\Sigma$ drops steeper with increasing $T$, as
illustrated in Fig.~\ref{fig:inverse_mc}.

It is nearly impossible to explain such an exotic pattern of
modifications on $\Sigma(T)$ from the dynamics of mass generation
alone (see Ref.~\cite{Fukushima:2012kc} for a scenario closed in the
chiral sector), and the coupling to confinement sector incorporated in
the PNJL model is still insufficient to realize this pattern.  In
other words, the finite-$T$ inverse magnetic catalysis is a serious
challenge to such model studies.  Clearly the PNJL model must miss
something in the confinement or gluonic sector.  In fact it has been
known from applications to high-density matter that the gluonic
potential used in the PNJL model lacks backreaction from quark
polarizations on gluon propagation (for the LLLA estimate for the
polarization effect, see Ref.~\cite{Fukushima:2011nu} and for the full
Landau level calculation, see
Refs.~\cite{Hattori:2012je,Hattori:2012ny}).  For another
nonperturbative backreaction, see Ref.~\cite{Bruckmann:2013oba}.
These missed diagrams
make the strong coupling $\alpha_s$ run with
energy~\cite{Ferrer:2014qka}, as $\lambda_k$ runs with $k$ in the
previous subsection.  The asymptotic freedom implies that $\alpha_s$
is a decreasing function of $B$ and $T$ both.  Some theoretical
calculations demonstrated that there may be a window in which $\Tc$
has a local minimum as a function of $B$, which explains the inverse
magnetic catalysis~\cite{Mueller:2015fka}.

Interestingly, in the vicinity of deconfinement transition, duality
between deconfined quark and confined hadronic degrees of freedom
could hold.  Therefore, the inverse magnetic catalysis may be
approached from the hadronic side.  In the hadronic phase there are
many composite states of quarks which carry electric charge such as
the charged pions and the proton.  Such charged hadrons with nonzero
spin become lighter significantly as a result of the Zeeman coupling
with the magnetic field.  In this way, as an extension from
Ref.~\cite{Endrodi:2013cs}, it has been numerically
confirmed that the ``critical'' temperature inferred from rapid changes
in thermodynamic quantities is shifted down toward a smaller value for
strong magnetic field~\cite{Fukushima:2016vix}.  This is a dual
picture to view the inverse magnetic catalysis.

\subsection{Toward more realistic descriptions}

We have idealized the physical setup assuming infinite volume and
spatial homogeneity, but in reality, the system size, that is the size
of matter distribution and/or applied magnetic field, is finite.  When
we discuss rotation effects later, it will be crucial to take the finite
size seriously; otherwise, the causality is violated.  One may think
that such finite size effects are anyway small corrections, but as we
will see here, qualitatively new physics is perceived from those
analyses.

\subsubsection{Solvable example of inhomogeneous magnetic field}
\label{sec:inhomogeneous}

It is generally a hard task to solve quantum field theory problems
without translational invariance.  Inhomogeneous electromagnetic
backgrounds break translational invariance, and there is no universal
algorithm to take account of such fields.  Thus, if any, some
theoretical exercises using analytically solvable examples would be
helpful for us to sharpen our feeling about the effect of
inhomogeneous fields.

One well-known solvable example is the Sauter-type potential.  The
Sauter potential literally means a time-dependent electric potential,
which was studied as an attempt to resolve the Klein
paradox~\cite{Sauter:1932gsa}, which we will discuss later in
Sec.~\ref{sec:electric}.  Here, with a magnetic counterpart of the
Sauter-type potential, the magnetic field is directed along the
$z$-axis and the spatial dependence is one-dimensional along the
$x$-axis like
\begin{equation}
  \bB(x) = \frac{B}{\cosh^2(k x)}\,\ez\,.
  \label{eq:SauterB}
\end{equation}
This magnetic distribution is peaked around $x\sim 0$ and the wave
number $k$ characterizes the typical scale of the spatial modulation.
The Sauter-type magnetic configuration can be described by the
following vector potential:
\begin{equation}
  A^2 = \frac{B}{k}\tanh(k x)\,,
\end{equation}
which smoothly reduces to $A^2=Bx$ in the limit of $k\to 0$.

The eigenvalue spectrum and the wave-functions can be found in
Ref.~\cite{Cangemi:1995ee} (in which the potential is called the
modified P\"{o}schl-Teller form) for both fermions and bosons.  We
will not repeat the derivation here but jump to the final results for
fermions.  The energy dispersion relation for Dirac particles under
$\bB(x)$ of Eq.~\eqref{eq:SauterB} is,
$\epsilon^2 = \lambda + p_z^2 + m^2$, where, for integer
$n\in[0,|\frac{1}{2}\pm \frac{eB}{k^2}|-\frac{1}{2}-\sqrt{\frac{p_y eB}{k^3}})$
(which also defines the range of $p_y$ to keep $n>0$) and
$\tiln=n+\frac{1}{2}$, the eigenvalue $\lambda$ is given as
  \begin{equation}
  \lambda_n^\pm = p_y^2 \biggl[ 1-\frac{(eB)^2}
    {(k^2\tiln - |k^2/2 \pm eB|)^2} \biggr] \mp eB
  - \biggl( k^2\tiln^2 -2\tiln \biggl|\frac{k^2}{2}
  \pm eB\biggr| + \frac{k^2}{4} \biggr)\,.
\end{equation}
This expression can become much simpler for small inhomogeneity if
only $eB > k^2/2$ is satisfied, and then we can remove the modulus to
simplify the above expression for $\lambda_n^\pm$ as
\begin{equation}
  \lambda_n^\pm = \biggl[ 1-\frac{p_y^2 k^2}{(eB-k^2 n^\pm)^2}
    \biggr] n^\pm\, (2eB - k^2 n^\pm )\,,
  \label{eq:spectrum_inhomo}
\end{equation}
with $n^+=n \in [0,\frac{eB}{k^2}-\sqrt{\frac{p_y eB}{k^3}})$
and $n^-=n+1 \in [1,\frac{eB}{k^2}-\sqrt{\frac{p_y eB}{k^3}})$,
which also constrains the possible range of $p_y$.  We can immediately
convince ourselves that the ordinary Landau quantization is recovered
for $k=0$.  Then, $\lambda_n^\pm=2eB n^\pm$ for $k=0$ and $n^\pm$
correspond to the spin up and down Landau levels.  In particular the
Landau zero mode, $n^+=0$, exists for one spin state only.  In this
case of $k=0$ the $p_y$-integration yields the Landau degeneracy
factor which is regularized by the system size.  In other words the
Landau degeneracy factor is proportional to the magnetic flux which
would diverge for infinitely large systems.

\begin{figure}
  \centering
  \includegraphics[width=0.4\textwidth]{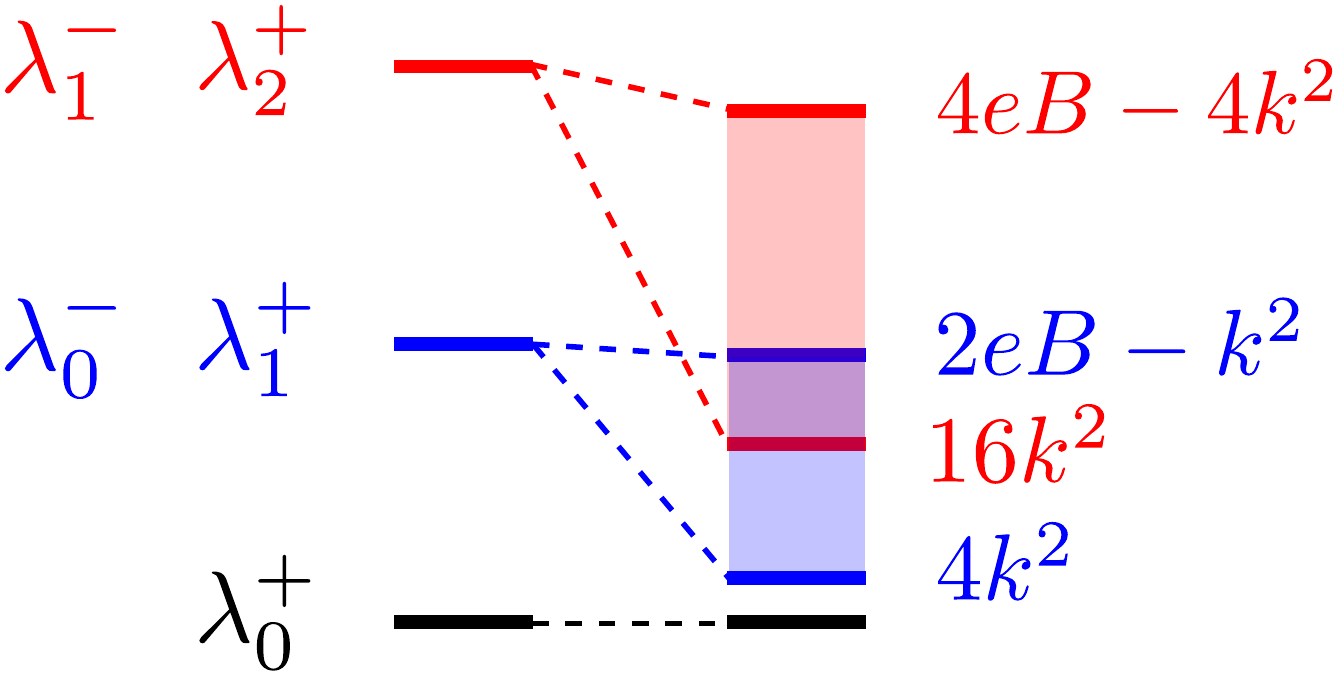}
  \caption{Schematic figure to show the eigenvalue spectra with and
    without spatial modulation of the magnetic field.  The left is the
  ordinary Landau quantized spectrum.  Each energy level is shifted
  with nonzero $k$ and the spectrum spreads due to the lack of
  degeneracy in $p_y$.  For subtlety about the lower bounds of
  spreading spectrum, see discussions in the text.}
  \label{fig:spectrum}
\end{figure}

The eigenvalue spectrum of Eq.~\eqref{eq:spectrum_inhomo} is very
useful for us to develop our intuition about the effect of
inhomogeneity.  The most notable feature is that, interestingly, even
though the magnetic field is inhomogeneous, the Landau zero mode,
$\lambda_0^+=0$, exists (but the degeneracy factor could be
modified).  This observation could be a manifestation of the
Atiyah-Singer index theorem; see Ref.~\cite{PhysRevA.19.2461} for more
discussions.  Moreover, the higher excited states are pushed down
overall by inhomogeneity.  Let us go a little more into concrete
numbers for the low-lying states.

For homogeneous magnetic field with $k=0$, the first Landau level is
located at $\lambda_1^+ = \lambda_0^- = 2eB$ with spin degeneracy.
The degeneracy with respect to $p_y$ is lost by $k\neq 0$.  Here we
shall assume $k^2\ll eB$ and expand the eigen-energies in terms of
$k^2/eB$.  We then approximately obtain $4k^2 \lesssim \lambda_1^+,
\lambda_0^- \lesssim 2eB-k^2$ up to the $k^2$ order.  Therefore, the
LLLA might be suspicious if $k^2$ is comparable to interested energy
scale even though $eB$ itself is sufficiently large, which is quite
natural because the magnetic field is then damped quickly.  In the same way,
the second Landau level is perturbed to spread over
$16k^2 \lesssim \lambda_2^+, \lambda_1^- \lesssim
4eB - 4k^2$, and so the minimal energy gap is again of order not $eB$
but $k^2$.  Therefore, inhomogeneous magnetic fields soften
one-particle excitation energies.  If $k^2\ll eB$, one can also
perform diagrammatic perturbative expansion, as done in
Ref.~\cite{Copinger:2016llk}, to get some confidence about the
generality of the above statement.

Now we should be cautious about the lower bounds in the $k\to 0$
limit.  Seemingly all lower bounds of spreading spectrum approach zero
(i.e. $4k^2\to 0$, $16k^2\to 0$, etc) as $k$ goes smaller, and if so,
the discrete spectrum at $k=0$ cannot be retrieved smoothly from the
$k\to 0$ limit.  In drawing Fig.~\ref{fig:spectrum} we assumed that
$p_y$ can be as large as $\sim 1/k$, but in reality, $p_y$ is cut off
by a combination of the magnetic field and the system size.  Then,
with $p_y$ bounded smaller than $1/k$ in such a way, the spectrum
shrinks to the discrete one in the $k\to 0$ limit as it should.

\subsubsection{Finite size and surface effect}
\label{sec:finite}

Instead of considering inhomogeneity in the applied magnetic field, we
will turn to a different but somehow related problem in this
subsection, that is, the finite size effect.  We do not want to break
translational invariance along the magnetic direction (i.e., the
$z$-axis), so the simplest geometrical setup is given by the
cylindrical coordinates, $(r,\theta,z)$.  Before considering the
boundary effect on magnetic systems, let us first take a close look at
the solutions of the free Dirac equation for $B=0$ in the cylindrical
coordinates.  For particles with positive energy, two helicity states
can be written down explicitly as
\begin{equation}
  u_+ = \frac{\rme^{-\rmi\omega t+\rmi p_z z}}{\sqrt{\omega+m}}
  \begin{pmatrix}
    (\omega+m) J_\ell(p_{\ell,k}r)\rme^{\rmi\ell\theta} \\
    0 \\
    p_z J_\ell(p_{\ell,k}r)\rme^{\rmi\ell\theta} \\
    \rmi p_{\ell,k} J_{\ell+1}(p_{\ell,k}r)\rme^{\rmi(\ell+1)\theta}
  \end{pmatrix}\,,\quad
  u_- = \frac{\rme^{-\rmi\omega t+\rmi p_z z}}{\sqrt{\omega+m}}
  \begin{pmatrix}
    0 \\
    (\omega+m) J_{\ell+1}(p_{\ell,k}r)\rme^{\rmi(\ell+1)\theta} \\
    -\rmi p_{\ell,k} J_\ell(p_{\ell,k}r)\rme^{\rmi \ell\theta} \\
    -p_z J_{\ell+1}(p_{\ell,k}r)\rme^{\rmi(\ell+1)\theta}
  \end{pmatrix}\,,
\end{equation}
where the onshell condition is $\omega^2=p_{\ell,k}^2+p_z^2+m^2$.  We
see that these $u_\pm$ are simple generalization from the standard
expression~\eqref{eq:spinors} with $\xi_\pm=(1,0)^T, (0,1)^T$ for
$u_\pm$.  There are, however, two major differences from
Eq.~\eqref{eq:spinors}.  One is that the transverse coordinates, $x$
and $y$ are replaced by $r$ and $\theta$ whose conjugate momenta are,
respectively, $p_{\ell,k}$ and $\ell$.  The other is that $p_{\ell,k}$
is discretized by the system size or the cylinder radius $R$.  That
is, no flux condition at $r=R$ leads to the discretized momenta as
\begin{equation}
  p_{\ell,k} = \begin{cases}
    \xi_{\ell,k} R^{-1} & \text{for}\quad \ell=0,1,\dots \\
    \xi_{-\ell-1,k} R^{-1} & \text{for}\quad \ell=-1,-2,\dots
  \end{cases}
  \label{eq:disc_p}
\end{equation}
where $\xi_{\ell,k}$ is the $k$-th zero of the Bessel function
$J_\ell(x)$.  We can immediately write down the
anti-particle solutions using the $\calC$-parity transformation, i.e.,
$v_\pm = \rmi\gamma^2 u_\pm^\ast$ as
\begin{equation}
  v_+ = \frac{\rme^{\rmi\omega t-\rmi p_z z}}{\sqrt{\omega+m}}
  \begin{pmatrix}
    -\rmi p_{\ell,k} J_{\ell+1}(p_{\ell,k}r)\rme^{-\rmi (\ell+1)\theta} \\
    -p_z J_\ell(p_{\ell,k}r)\rme^{-\rmi \ell\theta} \\
    0 \\
    (\omega+m) J_\ell(p_{\ell,k}r)\rme^{-\rmi \ell\theta}
  \end{pmatrix}\,,\quad
  v_- = \frac{\rme^{\rmi\omega t-\rmi p_z z}}{\sqrt{\omega+m}}
  \begin{pmatrix}
    -p_z J_{\ell+1}(p_{\ell,k}r)\rme^{-\rmi (\ell+1)\theta} \\
    -\rmi p_{\ell,k} J_\ell(p_{\ell,k}r)\rme^{-\rmi \ell\theta} \\
    -(\omega+m) J_{\ell+1}(p_{\ell,k}r)\rme^{-\rmi (\ell+1)\theta} \\
    0
  \end{pmatrix}\,.
\end{equation}
Again, it is clear that these are generalizations from
Eq.~\eqref{eq:spinors} with appropriate choice of $\eta_s$.  We also
note that $u_\pm$ have the angular momentum $j=\ell+1/2$ and $v_\pm$
have $-j$ along the $z$-axis.

Now, next, it is time to activate finite magnetic field.  The
positive-energy particle states with the $z$-component of the
angular momentum, $j=\ell+1/2$, are only slightly modified
as~\cite{Chen:2017xrj}
\begin{equation}
  \begin{split}
  &u_+ = \frac{\rme^{-\rmi\omega t+\rmi p_z z}}{\sqrt{\omega+m}}
  \begin{pmatrix}
    (\omega+m) \Phi_\ell(\lambda_{\ell,k},\frac{1}{2}eB r^2)
    \rme^{\rmi\ell\theta} \\
    0 \\
    p_z \Phi_\ell(\lambda_{\ell,k},\frac{1}{2}eB r^2)
    \rme^{\rmi\ell\theta} \\
    \rmi p_{\ell,k} \Phi_{\ell+1}(\lambda_{\ell,k}-1,
    \frac{1}{2}eB r^2)\rme^{\rmi(\ell+1)\theta}
  \end{pmatrix},\\
  &u_- = \frac{\rme^{-\rmi\omega t+\rmi p_z z}}{\sqrt{\omega+m}}
  \begin{pmatrix}
    0 \\
    (\omega+m) \Phi_{\ell+1}(\lambda_{\ell,k}-1,\frac{1}{2}eB r^2)
    \rme^{\rmi(\ell+1)\theta} \\
    -\rmi p_{\ell,k}\Phi_\ell(\lambda_{\ell,k},
    \frac{1}{2}eB r^2)\rme^{\rmi \ell\theta} \\
    -p_z \Phi_{\ell+1}(\lambda_{\ell,k}-1,\frac{1}{2}eB r^2)
    \rme^{\rmi(\ell+1)\theta}
  \end{pmatrix},
  \end{split}
\end{equation}
in which the Bessel function is upgraded to a more general special
function.  With the confluent hypergeometric function (Kummer's
function of first kind), this upgraded functions are defined as
\begin{align}
  \Phi_{\ell\ge 0}(\lambda,x) &\equiv \frac{1}{\Gamma(\ell+1)}
  \sqrt{\frac{\Gamma(\lambda+\ell+1)}{\Gamma(\lambda+1)}}
  x^{\ell/2} \rme^{-x/2}\, {_1F_1} (-\lambda,\ell+1,x)\,, \\
  \Phi_{\ell< 0}(\lambda,x) &\equiv \frac{(-1)^{-\ell+1}}{\Gamma(-\ell+1)}
  \sqrt{\frac{\Gamma(\lambda+1)}{\Gamma(\lambda+\ell+1)}}
  x^{-\ell/2} \rme^{-x/2}\, {_1F_1} (-\lambda-\ell,-\ell+1,x)\,.
\end{align}
The discretization condition for the transverse momenta is also
changed as
$p_{\ell,k}=\sqrt{2eB\lambda_{\ell,k}}$ and
\begin{equation}
  \lambda_{\ell,k} = \begin{cases}
    \tilde{\xi}_{\ell,k} & \text{for}\quad \ell=0,1,\dots\\
    \tilde{\xi}_{-\ell-1,k}-\ell & \text{for}\quad
    \ell=-1,-2,\dots
  \end{cases}
  \label{eq:disc_B}
\end{equation}
where $\tilde{\xi}_{\ell,k}$ is $k$-th zero of
${_1 F_1}(-\xi,\ell+1,\alpha)$ with $\alpha\equiv\frac{1}{2}eB R^2$.
We note that $\alpha$ is nothing but the Landau degeneracy factor,
$(eB/2\pi)\cdot \pi R^2$.  It would be more understandable if we see
Eq.~\eqref{eq:disc_B}, not as a finite-$B$ generalization of
Eq.~\eqref{eq:disc_p}, but as a finite-$R$ generalization of the
standard Landau levels.  In fact, if $R$ or $\alpha$ is sufficiently
large, ${_1 F_1}(-\xi,\ell+1,\alpha\ll 1)$ has $k$-th zero at
$\xi=k-1$ for nonnegative integer $\ell$.

\begin{figure}
  \centering
  \includegraphics[width=0.4\textwidth]{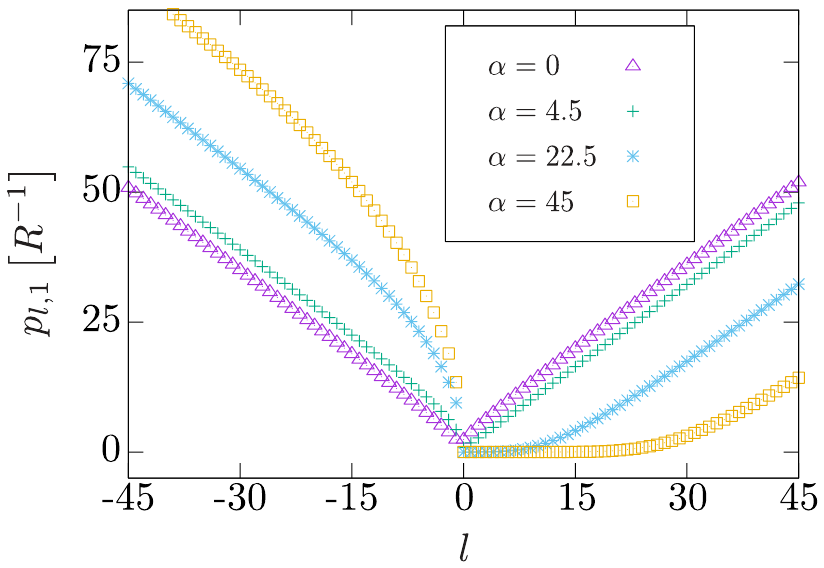}
  \caption{Lowest transverse momentum $p_{\ell,1}$ as a function of
    $\ell$ for various $\alpha=\frac{1}{2}eBR^2$.  Figure taken from
    Ref.~\cite{Chen:2017xrj}.}
  \label{fig:bd_deg}
\end{figure}

Figure~\ref{fig:bd_deg} shows $p_{\ell,1}$ as a function of $\ell$ for
various $\alpha$.  For zero magnetic field (i.e., $\alpha=0$), the
symmetry under $\ell\leftrightarrow -\ell-1$ is manifest.  In our
convention $j=\ell+\frac{1}{2}$ is the $z$ component of the total
angular momentum, and $\ell\leftrightarrow -\ell-1$ corresponds to
$j\leftrightarrow -j$.  Such symmetry of sign flip of the total
angular momentum is explicitly broken by the Zeeman energy from the
orbital-magnetic and spin-magnetic coupling.  We actually know that
the Landau zero mode has only one spin state;  as seen in
Fig.~\ref{fig:bd_deg}, for finite $\alpha$, nearly zero degenerate
states of $p_{\ell,1}$ spread only in the positive-$\ell$ side.  If
$B$ is strong enough and/or $R$ is large enough, the boundary effects
should be negligible and the Landau degeneracy of $p_{\ell,1}$ should
approach $\alpha$.  In Fig.~\ref{fig:bd_deg}, for $\alpha=45$ for
example, the flat bottom region ends around $\ell\lesssim 30$, and so
$\alpha$ is not yet large enough to enter the regime of the full
Landau degeneracy $\sim \alpha$.

One might think that the anti-particles could be constructed by
$v_\pm=\rmi \gamma^2 u_\pm^\ast$ as we did previously, but this
transformation does not work.  In the presence of external magnetic
field or vector potential in general, the $\calC$-parity symmetry is
explicitly broken (noting that the vector potentials are
$\calC$-odd), so $\rmi \gamma^2 u_\pm^\ast$ does not satisfy the Dirac
equation.  The direct calculations result in
\begin{equation}
  \begin{split}
  &v_+ = \frac{\rme^{\rmi\omega t-\rmi p_z z}}{\sqrt{\omega+m}}
  \begin{pmatrix}
    -\rmi p_{-\ell-1,k} \Phi_{-\ell-1}(\lambda_{-\ell-1,k},\frac{1}{2}eB r^2)
    \rme^{-\rmi (\ell+1)\theta} \\
    -p_z \Phi_{-\ell}(\lambda_{-\ell-1,k}-1,\frac{1}{2}eB r^2)
    \rme^{-\rmi \ell\theta} \\
    0 \\
    (\omega+m) \Phi_{-\ell}(\lambda_{-\ell-1,k}-1,\frac{1}{2}eB r^2)
    \rme^{-\rmi \ell\theta}
  \end{pmatrix},\\
  &v_- = \frac{\rme^{\rmi\omega t-\rmi p_z z}}{\sqrt{\omega+m}}
  \begin{pmatrix}
    -p_z \Phi_{-\ell-1}(\lambda_{-\ell-1,k},\frac{1}{2}eB r^2)
    \rme^{-\rmi (\ell+1)\theta} \\
    -\rmi p_{-\ell-1,k} \Phi_{-\ell}(\lambda_{-\ell-1,k}-1,\frac{1}{2}eB r^2)
    \rme^{-\rmi \ell\theta} \\
    -(\omega+m) \Phi_{-\ell-1}(\lambda_{-\ell-1,k},\frac{1}{2}eB r^2)
    \rme^{-\rmi (\ell+1)\theta} \\
    0
  \end{pmatrix}\,.
  \end{split}
\end{equation}
We note that in Ref.~\cite{Chen:2017xrj} the anti-particle solutions
are given with $-\ell-1$ changed to $\ell$, which means that the above
states have the angular momentum $-j=-\ell-1/2$, while such
anti-particle states in Ref.~\cite{Chen:2017xrj} have
$-(-\ell-1)-1/2=j$.  The convention of Ref.~\cite{Chen:2017xrj} is
practically useful, but the spin quantum number in the above
expression is more in accord to our intuition.  In the present choice
the approximate Landau zero modes of $v_\pm$ appear for
negative-$\ell$ region since above $v_\pm$ involve $p_{-\ell-1,k}$.
Therefore, the angular momenta $j$ of those nearly Landau zero modes
should be positive.  In this way, we observe a peculiar pattern in the
Landau zero modes that, for particles and anti-particles, the favored
spin alignments are opposite and yet the total angular momenta are
degenerated.  This corresponds to the fact that the magnetic responses
of particles and anti-particles should be opposite, but if a small
rotation is introduced perturbatively, the rotational effect would not
discriminate particles and anti-particles.

Now that a complete set of wave-functions comes by, we can compute the
propagator and any operator expectation values perturbatively.  In
fact, in Ref.~\cite{Chen:2017xrj}, the scalar condensate has been
calculated in the local density approximation.  Then, it has been
found that the magnetic catalysis is significantly strengthened near
the boundary surface, $r\sim R$, due to the accumulation of
large-$\ell$ wave-functions.  Anomalous contributions near the surface
may play a crucial role to fulfill global conservation of some
physical charges.

Summarizing the discussions in this subsection, we constructed
explicit solutions of the Dirac equation with a boundary condition at
$r=R$.  The finiteness of the system is important for phenomenological
applications;  the Landau degeneracy factor diverges for infinite
sized systems, and such a simple estimate, $(eB/2\pi)\cdot \pi R^2$,
based on homogeneous treatment is not necessarily a good
approximation.  Moreover, the identification of anti-particles is
nontrivial.  It was not my intention to bother readers with technical
details, but above expressions are firm bases to visualize the
physics.

\section{Dynamical Problems with Electromagnetic Fields}
\label{sec:electric}

We will explore some formalisms to cope with not only the magnetic
field but also the electric field.  We will start with a case with
only the electric field to study the Schwinger Mechanism, and then
turn on a constant magnetic field.  The coexistence of the parallel
electromagnetic fields would break the $\calP$- and
$\calC\calP$-symmetries, and such an electromagnetic configuration is
a theoretically idealized setup to probe the chiral anomaly.  We will
introduce an idea called the chiral magnetic effect (CME) as a clear
signature for the chiral anomaly.

\subsection{Schwinger Mechanism}
\label{sec:schwinger}

Quantum field theory calculus was completed in
Ref.~\cite{PhysRev.82.664} and the generating functional, which is an
amplitude from the past vacuum to the future vacuum, was found to
acquire an imaginary part in the presence of constant electric field.
The appearance of imaginary part generally signifies a kind of
instability.  There have been some confusions about the theoretical
interpretation about the imaginary part (see Ref.~\cite{Cohen:2008wz}
for judicious discussions on this issue).  For some historical
backgrounds together with the treatment of the Sauter-type potential,
see also Ref.~\cite{Dunne:2004nc}.  A comprehensive review on the
Schwinger Mechanism including latest developments can be found in
Ref.~\cite{Gelis:2015kya}.  For beginners a
textbook~\cite{itzykson1985quantum} is also recommendable.  In the
present review I would not touch such a subtle argument, but I would
rather prefer to formulate the same physics in terms of the particle
production amplitude~\cite{Fukushima:2009er}.

The particle production can occur whenever the energy-momentum
conservation is satisfied and the quantum numbers are matched.  In the
case of the pair production, the momentum conservation can be
satisfied if only the emitted particle and anti-particle are
back-to-back placed.  Because such particle and
anti-particle carry finite energy, the external fields should inject
an energy into the system.  It is obviously impossible to balance the
energy with homogeneous magnetic field only which gives no work on
charge particles.  If a pulse-like electromagnetic field externally
disturbs the system, a virtual (offshell) photon from the electric
field can decay into a pair of a particle and an anti-particle, and in
this case, a vertex of photon, particle, and anti-particle gives a
tree-level contribution to the pair production amplitude.  Thus,
nothing in particular is special in the pair production process with
pulse-like fields.

A surprise comes from the fact that even a constant electric field can
supply a finite energy, which allows for the pair production.  As we
will discuss later, the introduction of constant electric field
assumes a time-dependent background vector potential.  The
time-dependence is, however, infinitesimal, and thus, the energy
transfer from a perturbative process with this background field is
infinitesimal.  Therefore, a single scattering cannot meet the energy
conservation law, but then, what about multiple scatterings?  Even if
the energy transfer from each scattering is infinitesimal, infinitely
many scatterings eventually amount to a finite energy so that the
energy conservation can be satisfied.  This nonperturbative process
for the pair production driven by constant electric field is called
the Schwinger Mechanism, i.e.,

\[
\framebox{
  \begin{minipage}{0.8\textwidth}
    Schwinger Mechanism: The vacuum under constant electric field is
    unstable to produce pairs of a particle and an anti-particle via
    nonperturbative processes.
  \end{minipage}
}
\]

It is often said that the Schwinger Mechanism is a result of quantum
tunneling.  Actually, we can formulate a pair production as a
conversion process from an anti-particle (negative-energy state) in
the Dirac sea to a particle (positive-energy state).  The particle and
the anti-particle states are gapped by the particle mass, $m$, which
may well be regarded as a sort of activation energy, and then the
electric field is like the temperature.  In this way there have been
some theoretical speculations about a possible connection between the
Schwinger Mechanism and thermal nature of produced particles.  In
fact, as we will elaborate below, the tunneling amplitude is
characterized by the Bogoliubov coefficient in a way very similar to
the Hawking radiation process from the blackhole (see
Ref.~\cite{tHooft:1996rdg} for detailed derivation of the radiated
spectrum).  There is a significant difference, however, in the final
expressions;  the rate in the Schwinger Mechanism is exponentially
suppressed like $\sim \exp[-\pi m^2/(eE)]$ which is much smaller than
the Boltzmann factor $\sim e^{-m/T}$ for large $m$.

\subsubsection{Solvable example of time-dependent electric fields}

\begin{figure}
  \centering
  \includegraphics[width=0.6\textwidth]{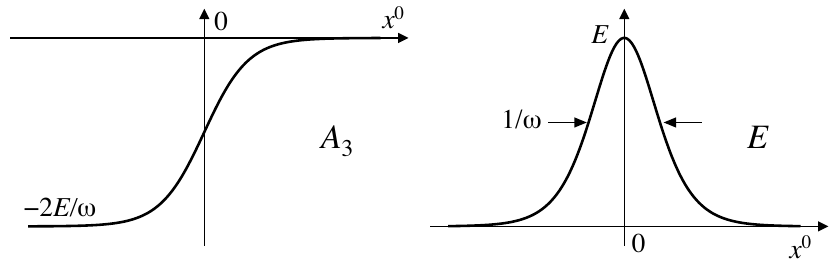}
  \caption{Profile of the vector potential and the electric field as
    functions of $t$.  Figures taken from
    Ref.~\cite{Fukushima:2009er}.}
  \label{fig:E_profile}
\end{figure}

Here, I would not intend to explain mathematical techniques to solve
the differential equation, but using a solvable example I would
present a demonstration to concretize some important concepts on the
Schwinger Mechanism.  The solvable example, i.e., the Sauter-type
electric profile is
\begin{equation}
  \bE(t) = \frac{E}{\cosh^2(\omega t)} \ez\,,
  \label{eq:Sauter_E}
\end{equation}
which is realized by the following vector potential,
\begin{equation}
  A_3(t) = \frac{E}{\omega} \bigl[ \tanh(\omega t) - 1 \bigr]\,.
  \label{eq:Sauter_A}
\end{equation}
This example was found by Sauter in 1932~\cite{Sauter:1932gsa} to
discuss the Klein paradox, that is a puzzle of inevitable transmission
of fermions over electric barriers.  As a matter of fact, the
Schwinger Mechanism is a field-theoretical solution to the Klein
paradox.  Coming back to the calculation, we notice that $A_3$ in
Eq.~\eqref{eq:Sauter_A} takes different values in the in-state at
$t\to-\infty$ and the out-state at $t\to+\infty$, while $\bE(t)$ in
Eq.~\eqref{eq:Sauter_E} goes to zero at both $t\to\pm\infty$, as
displayed in Fig.~\ref{fig:E_profile}.  The background gauge potential
shifts the energy dispersion relations of the in- and the out-states,
respectively, as
\begin{equation}
  E^{\rm (in)} = \sqrt{(p_3-2\lambda\omega)^2+p_\perp^2+m^2}\,,\qquad
  E^{\rm (out)} = \sqrt{p_3^2 + p_\perp^2 + m^2}\,,
\end{equation}
where $\lambda\equiv eE/\omega^2$ is defined.

\begin{figure}
  \centering
  \includegraphics[width=0.5\textwidth]{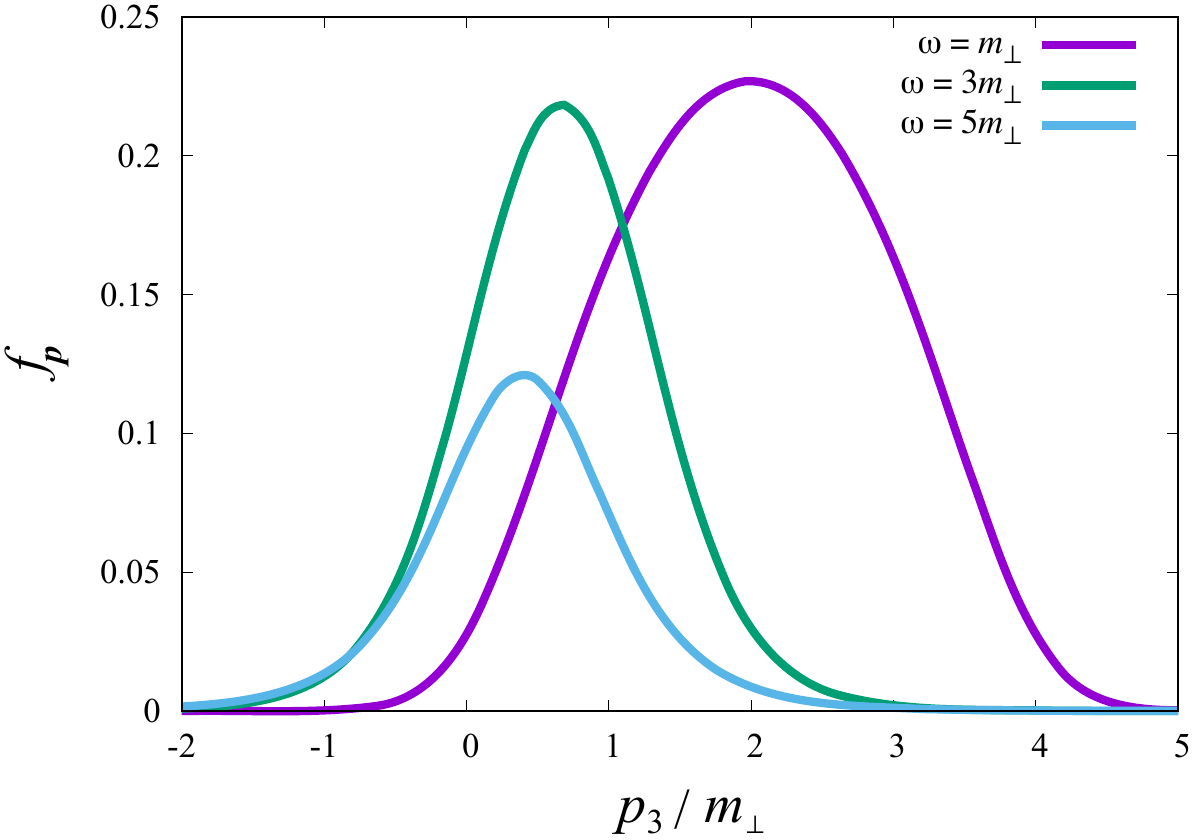}
  \caption{Produced particle spectrum for $eE=2m_\perp^2$ and various
    pulse parameter $\omega$.  The distribution, $f_{\bp}$, is
    normalized to be a dimensionless density, such that its phase
    space integral, $V\int\frac{\rmd^3\bp}{(2\pi)^3}\,f_{\bp}$, should
    be a dimensionless number.}
  \label{fig:E_production}
\end{figure}

We can quantify the particle production by computing the Bogoliubov
coefficients, that is, coefficients connecting the
creation/annihilation operators in the in- and the out-states.  For
this purpose, we can solve the equation of motion from the initial
condition of either a particle state or an anti-particle state, that is,
\begin{equation}
  \psi_{\bp}^{(1)}(t\sim -\infty) \to u(\bp)\,\rme^{-\rmi E^{\rm (in)}t}\,,\qquad
  \psi_{\bp}^{(2)}(t\sim -\infty) \to v(\bp)\,\rme^{\rmi E^{\rm (in)}t}\,,
\end{equation}
which evolves into
\begin{equation}
  \begin{split}
  &\psi_{\bp}^{(1)}(t\sim +\infty) \to
  A_{\bp}\,u(\bp)\,\rme^{-\rmi E^{\rm (out)}t}
  - B_{-\bp}^\ast\,v(-\bp)\,\rme^{\rmi E^{\rm (out)}t}\,,\\
  &\psi_{\bp}^{(2)}(t\sim +\infty) \to
  A_{\bp}^\ast\,v(\bp)\,\rme^{\rmi E^{\rm (out)}t}
  + B_{-\bp}\,u(-\bp)\,\rme^{-\rmi E^{\rm (out)}t}\,.
  \end{split}
\end{equation}
If no electric effect is applied, there is no mixing between particles
and anti-particles and thus $A_{\bp}=1$ and $B_{\bp}=0$ trivially.  A nonzero
$B_{\bp}$ represents the tunneling amplitude from the anti-particle to
the particle (and vice versa).  With these coefficients for fields,
taking account of the field expression in terms of
creation/annihilation operators, we can identify the Bogoliubov
coefficients as
\begin{equation}
  \alpha_{\bp} = \sqrt{\frac{E^{\rm (out)}}{E^{\rm (in)}}} A_{\bp}\,,\qquad
  \beta_{\bp} = \sqrt{\frac{E^{\rm (out)}}{E^{\rm (in)}}} B_{\bp}\,.
\end{equation}
The produced particle distribution, $f_{\bp}$, which is defined by the
number operator expectation value as
$f_{\bp}\equiv V^{-1}\langle{\rm in}|a_{{\rm out},\bp}^\dag a_{{\rm
    out},\bp}|{\rm in}\rangle$, is given by
\begin{equation}
  f_{\bp} = |\beta_{\bp}|^2\,.
\end{equation}
For the Sauter-type potential the solution is known and the Bogoliubov
coefficient reads~\cite{Nikishov:1970br}:
\begin{equation}
  f_{\bp} = \frac{\sinh[\pi(\lambda-\mu+\nu)]\sinh[\pi(\lambda+\mu-\nu)]}
  {\sinh(2\pi\mu)\sinh(2\pi\nu)}\,,
\end{equation}
where $\mu\equiv E^{\rm (in)}/(2\omega)$ and
$\nu\equiv E^{\rm (out)}/(2\omega)$.  This distribution function is
plotted for $eE=2 m_\perp^2$ [where $m_\perp^2\equiv 2(p_1^2+p_2^2+m^2)$]
as a function of $p_3$ in Fig.~\ref{fig:E_production}.

From Fig.~\ref{fig:E_production} we see that the pair production rate
is peaked at $p_3=eE/\omega$ for which the above expression becomes
much simpler with $E^{\rm (in)}=E^{\rm (out)}$ and thus $\mu=\nu$.  At
this peak position, we can take $\omega\to 0$ limit to consider
constant electric field, to find that the above expression reduces to
\begin{equation}
  f_{p_3=eE/\omega} \;\;\underset{\omega\to 0}{\longrightarrow}\;\;
  \exp\biggl(-\frac{\pi m_\perp^2}{eE} \biggr)\,.
\end{equation}
This exponential factor is typical for the Schwinger Mechanism.  The
transverse momentum integration gives a prefactor by the phase space
$\propto eE$.  The remaining exponential characterizes the critical
value, sometimes called the Schwinger critical electric field, whose
expression is
\begin{equation}
  (eE)_{\text{critical}} \sim \pi m^2\,.
\end{equation}
I must emphasize that nothing is truly critical at this critical
electric field.  There is no phase transition nor avalanche phenomenon
but the above relation is just an order estimate for the electric
field necessary for a sizable amount of pair production.

\subsubsection{Worldline instanton approximation}

Now we would like to turn on a magnetic field on top of electric
field.  If these fields are constant, such a problem is solvable,
though the technical details are quite complicated (see
Ref.~\cite{Dunne:2004nc}).  Here, we shall take a bypass, i.e.,
instead of getting a rigorous answer, we make less efforts and
approximate formulas correctly on the qualitative level.  We
further assume $\bB\parallel \bE$ in this section.  In later
discussions we will be interested in the chiral anomaly, which is
detectable with background $\bB\cdot\bE\neq 0$.

In this review I will entirely neglect backreaction from gauge
fluctuations (see Ref.~\cite{Tanji:2008ku} for simulations with
backreaction), but focus on the response of matter under fixed gauge
background.  Thus, it is sufficient to consider a fermionic part of
the action.  Because fermions enter gauge theories in a quadratic form
unlike the NJL model where four fermionic interactions are assumed, we
can simply identify the Dirac determinant as the full effective
action, that is,
\begin{equation}
  \Gamma[A] = -\rmi \ln\det\bigl[ \rmi\feyn{D} - m\bigr]
  = -\frac{\rmi}{2}\ln\det\Bigl[D^2 + m^2
    + \frac{\rmi e}{2}\sigma\cdot F \Bigr]\,,
\label{eq:fermion_action}
\end{equation}
where the covariant derivative is $D\equiv \partial+\rmi eA$.
From the middle to the right expression we multiplied
$\gamma_5[\rmi\feyn{D}-m]\gamma_5=-\rmi\feyn{D}-m$ which duplicates
the Dirac determinant.  We note that this trick is useful as long as
the $\gamma_5$-Hermiticity holds;  for example, a finite chemical
potential breaks this property.  The last term involving the spin tensor,
$\sigma^{\mu\nu}\equiv \frac{1}{2}[\gamma^\mu,\gamma^\nu]$, represents
the spin-magnetic coupling.


It is straightforward to reexpress $\Gamma[A]$ in terms of the
proper-time integration, as we saw in Eq.~\eqref{eq:S0proper}, using a
general operator relation,
\begin{equation}
  -\tr \ln \biggl(\frac{\calA}{\calB} \biggr)
  = \int_0^\infty \rmd s\, s^{-1}\,\tr \bigl(\rme^{-\calA s}
  - \rme^{-\calB s} \bigr)\,.
\end{equation}
Here, $\calA$ is to be replaced by the Dirac operator, and
$\rme^{-\calA s}$ is then regarded as time evolution by a
``Hamiltonian'', $\calA$, with an ``imaginary-time'' $s$, which can be
viewed as a quantum mechanical system evolving from $x_\mu$ at $\tau=0$
to the same $x_\mu$ at $\tau=s$ due to the trace nature.  We can
adopt Feynman's path integral representation to describe this quantum
mechanical evolution for $x_\mu(\tau)$.  For step-by-step
transformations in detail, see Ref.~\cite{Schubert:2001he}.  After
all, the derivatives or the momentum variables in the Hamiltonian are
translated into $\rmd x/\rmd \tau$ in the Lagrangian system as
\begin{equation}
  \Gamma[A] = \frac{1}{2} \int_0^\infty \rmd s\,s^{-1}\, \rme^{-m^2 s}
  \oint \calD x_\mu \,\exp\biggl\{
  -\int_0^s \rmd\tau \biggl[\frac{1}{4}\biggl(\frac{\rmd x}{\rmd \tau}
    \biggr)^2 + \rmi eA\cdot \frac{\rmd x}{\rmd \tau} \biggr]
  \biggr\} \Phi[A]\,,
\end{equation}
where $\oint$ represents the integration under the periodic boundary
condition; $x_\mu(\tau=0)=x_\mu(\tau=s)$.  It should be mentioned that
the above expression assumes the Wick rotation to Euclidean variables,
$(x_1,x_2,x_3,x_4)$.  The last part, $\Phi[A]$, corresponds to the
matrix part in Eq.~\eqref{eq:fermion_action}.  For our special problem
with $\bE=E\ez$ and $\bB=B\ez$, this last part can become factorized
into the electric and the magnetic contributions as
\begin{equation}
  \Phi[A] = \tr\calP \exp\biggl(\frac{\rmi e}{2}\int_0^s
  \rmd\tau\, \sigma\cdot F\biggr)
  = 4\cos\biggl(\int_0^s \rmd\tau\,eE\biggr)\,
  \cosh\biggl(\int_0^s \rmd\tau\,eB\biggr)\,.
\end{equation}
Therefore, the fermionic effective action takes the following
semi-factorized form:
\begin{equation}
  \Gamma[A] = 2\int_0^\infty \rmd s\,s^{-1}\,\rme^{-m^2 s}
  \calK_E\,\calK_B\,.
\end{equation}
Here, introducing a dimensionless variable $u=\tau/s$, we can write
down the explicit forms of the electric and the magnetic terms as
\begin{align}
  \calK_E &= \oint \calD x_3\,\calD x_4\,\cos\biggl(s\int_0^1 \rmd u\,
  eE \biggr)\,
  \exp\biggl[ -\int_0^1 \rmd u\biggl(\frac{\dot{x}_3^2+\dot{x}_4^2}
    {4s} + \rmi e A_3 \dot{x}_3 + \rmi e A_4 \dot{x}_4\biggr)
    \biggr]\,,\\
  \calK_B &= \oint \calD x_1\,\calD x_2\,\cosh\biggl(s\int_0^1 \rmd u\,
  eB \biggr)\,
  \exp\biggl[ -\int_0^1 \rmd u\biggl(\frac{\dot{x}_1^2+\dot{x}_2^2}
    {4s} + \rmi e A_1 \dot{x}_1 + \rmi e A_2 \dot{x}_2\biggr)
    \biggr]\,,
\end{align}
where $\dot{x}_\mu$ represents the $u$-derivative of $x_\mu$.
Interestingly, for constant electromagnetic fields, these kernels can
be evaluated without approximation, that is,
\begin{equation}
  \calK_B = \frac{eB}{4\pi}\,\coth(eB s)\,,\qquad
  \calK_E = \frac{eE}{4\pi}\,\cot(eE s)\,.
\end{equation}
Then, the effective action can be expressed without approximation as
\begin{equation}
  \Gamma = \frac{e^2 EB}{8\pi^2}\int_{1/\Lambda^2}^\infty
  \rmd s\,s^{-1}\,\coth(eB s)\,\cot(eE s)\,\rme^{-m^2 s}\,.
\end{equation}
This proper-time integration could be directly performed, which is
dominated by pole contributions around $s=n\pi/(eE)$ where
$\cot(eE s)$ is singular.  This singularity structure implies that
$\Gamma$ must have a series expansion like
$\sim\sum_n c_n \coth(n\pi B/E)\rme^{-n\pi m^2/eE}$ with some
coefficients $c_n$ of order unity.

An exact solution is of course useful, but I would rather prefer a
more adaptive method applicable to a wide variety of physics problems.
The worldline instanton approximation is such a flexible strategy and
at the same time some interesting physical interpretation is possible.
To see this, let us go back to the expression of $\calK_E$ in terms of
$x_3$ and $x_4$.  Then, before performing the $x_3$ and $x_4$
integrations, we shall first treat the $s$-integration.  There are two
completing $s$ dependence on the exponential;  one is from the mass,
$-m^2 s$, which favors smaller $s$, and the other is from the electric
kinetic term, $-\int \rmd u (\dot{x}_3^2+\dot{x}_4^2)/(4s)$, which
favors larger $s$.  One might have thought that the magnetic sector
has a similar term, but we already know the answer after the
integration, which behaves like $\coth(eB s)\sim 1$ for large $s$, and
has no effect on the location of the saddle point.

We follow the physics arguments in Ref.~\cite{Schutzhold:2008pz}.  We
can find the saddle-point to approximate the $s$-integration at
\begin{equation}
  s^\ast = \frac{1}{2m}\sqrt{\int_0^1 \rmd u\,
    (\dot{x}_3^2 + \dot{x}_4^2)}\,.
\end{equation}
Then, the effective action is approximated as
\begin{equation}
  \Gamma\approx \frac{2}{m}\oint \calD x_3\,\calD x_4\,
  \sqrt{\frac{\pi}{s^\ast}}\,\cos(eE s^\ast)\,\rme^{-S(x_3,x_4)}\,
  \calK_B(s^\ast)\,,
\end{equation}
where we can easily get $S(x_3,x_4)$ on the exponential from this
expression of $s^\ast$, that is,
\begin{equation}
  S(x_3,x_4) = m\sqrt{\int_0^1\rmd u\,(\dot{x}_3^2 + \dot{x}_4^2)}
  + \int_0^1 \rmd u\,(\rmi e A_3 \dot{x}_3 + \rmi e A_4\dot{x}_4)\,.
\end{equation}
We should then take care of the $x_3$ and $x_4$ integrations.  If $m$
is large, which is normally the case for the Schwinger problem, we can
utilize a semi-classical approximation, i.e., the $x_3$ and $x_4$
integrations should be dominated by classical trajectories which
(locally) minimize $S(x_3,x_4)$.  The minimization condition is
nothing but the equation of motion, and the solutions of the equation
of motion are commonly referred to as ``worldline instantons'' for the
reason explained later.  It is very easy to take a variation on
$S(x_3,x_4)$ to find the equation of motion as
\begin{equation}
  \frac{m\ddot{x}_i}{\displaystyle \sqrt{\int \rmd u\,(\dot{x}_3^2
      +\dot{x}_4^2)}} = \rmi e F_{ij}\, \dot{x}_j
  \label{eq:world_eom}
\end{equation}
for $i,j=3,4$.  A proper combination of these equations immediately
proves $\dot{x}_3^2+\dot{x}_4^2=\text{(const)}$, which simplifies the
denominator in the left-hand side of the equation of motion.  Plugging
$F_{34}=\rmi E$ (where $F_{34}$ is the Euclidean field strength and
$E$ is the Minkowskian physical electric field), imposing the periodic
boundary condition $x_i(u=0)=x_i(u=1)$, and choosing the initial
condition $x_4(u=0)=\dot{x}_3(u=0)=0$, we can obtain the solutions of
the equation of motion:
\begin{equation}
  \bar{x}_3(u) = \frac{m}{eE}\cos(2\pi n u)\,,\qquad
  \bar{x}_4(u) = \frac{m}{eE}\sin(2\pi n u)\,,
\end{equation}
characterized by $n=1,2,\dots \in \mathbb{Z}^+$.  These are very
interesting solutions;  $\bar{x}_3^2+\bar{x}_4^2=(m/eE)^2$ in
Euclidean spacetime implies $\bar{x}_3^2 - \bar{x}_0^2 = (m/eE)^2$ in
Minkowskian spacetime making hyperbolic trajectories, which are
solutions under an acceleration.
With these solutions, we see that the saddle point of $s$ is located
at $s_n^\ast = n\pi/(eE)$, which coincides with the singularities in
$\calK_E$.  Thus, we arrive at the same conclusion of the integration
dominated around $s\sim s_n^\ast$ though we exchanged the order of the
integrations;  $s$ first and $x_{3,4}$ later or vice versa.  The
action with these solutions becomes:
\begin{equation}
  \bar{S} = S[\bar{x}_3(u),\bar{x}_4(u)]
  = \frac{n\pi m^2}{eE}\,.
\end{equation}
The prefactor in the saddle point approximation can be also
calculated, and then we eventually reach the single pair ($n=1$)
production rate (i.e., pair numbers per unit volume and time) inferred
from $w = 2\Im\Gamma_{\rm M}|_{n=1}/V=-2\Re\Gamma_{\rm E}|_{n=1}/V$. 
Such an expression for the particle production rate in the presence of
parallel electromagnetic fields will be, in the next subsection, the key
equation for our analysis on the chiral anomaly.  What we learnt here
is summarized as follows:
\[
\framebox{
  \begin{minipage}{0.8\textwidth}
    Schwinger Pair Production Rate:
    The rate of single pair production in the presence of paralle
    electric and magnetic fields is given by the formula:
    \begin{equation}
      w = \frac{e^2 EB}{4\pi^2}\coth\biggl(\frac{\pi B}{E}
      \biggr)\,\exp\biggl(-\frac{\pi m^2}{eE}\biggr)\,.
      \label{eq:Schwinger}
    \end{equation}
  \end{minipage}
}
\]

\begin{figure}
  \centering
  \includegraphics[width=0.4\textwidth]{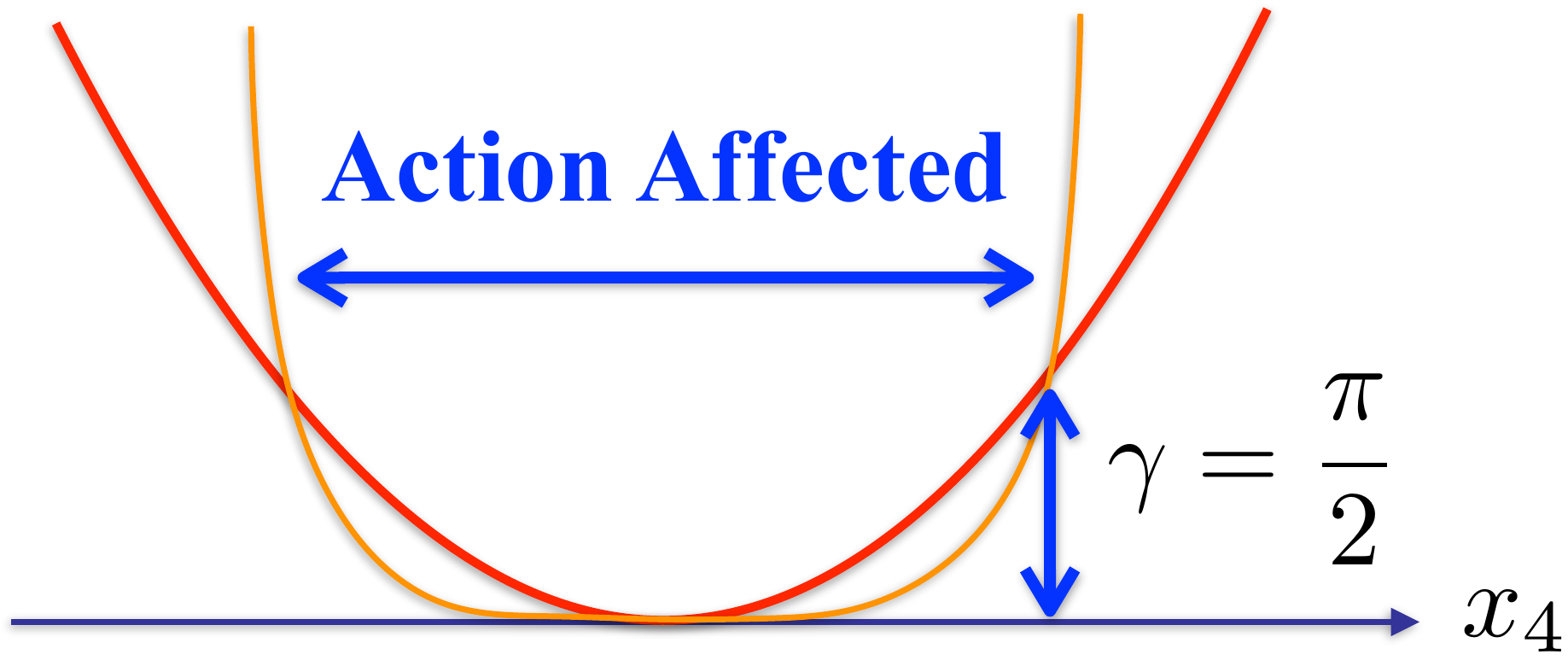}
  \caption{Schematic picture to understand the dynamically assisted
    Schwinger Mechanism.}
  \label{fig:assisted}
\end{figure}

One notable advantage in this method of the worldline instanton
approximation is the generality which can be directly applied to a
case with time-dependent electric perturbation.  A very interesting
idea has been proposed in Ref.~\cite{Schutzhold:2008pz};  the
Sauter-type potential is weakly perturbed from Eq.~\eqref{eq:Sauter_E}
as follows:
\begin{equation}
  \bE(t) = \frac{E}{\cosh^2(\Omega t)} \ez
  + \frac{\varepsilon}{\cosh^2(\omega t)} \ez\,,
  \label{eq:assisted_E}
\end{equation}
where $\omega\gg\Omega$.  The corresponding vector potential in terms
of Euclidean time $x_4$, up to irrelevant constants, reads:
\begin{equation}
  A_3(x_4) = -\rmi\frac{E}{\Omega}\tan(\Omega x_4)
  - \rmi \frac{\varepsilon}{\omega}\tan(\omega x_4)\,.
\end{equation}
In the worldline instanton approximation, we can make use of our
knowledge on classical mechanics.  Even without solving the equation
of motion, in classical mechanics, the energy conservation law easily
derived from the equation of motion can give us an intuitive picture
about the motion.  In fact, the equation of
motion~\eqref{eq:world_eom} leads to the conservation law for the sum
of the kinetic and the potential energies as
\begin{equation}
  \dot{x}_4^2 - \biggl(\frac{2\pi n}{eE}\biggr)^2\bigl[
    eA_3(x_4)\bigr]^2 = \text{(const.)}
\end{equation}
For sufficiently small $\Omega$, the first term in $A_3(x_4)$ is well
approximated by $E x_4$ describing a homogeneous electric field.  The
second term, $(\varepsilon/\omega)\tan(\omega x_4)$ is infinitesimal
for small $\varepsilon$ except when $\omega x_4$ approaches $\pi/2$
where $\tan(\omega x_4)$ diverges.  Therefore, in this case,
$[eA_3(x_4)]^2$ is an energy potential given by a superposition of a
quadratic term $\propto x_4^2$ and infinite energy barriers at
$\omega x_4=\pm\pi/2$ as sketched in Fig.~\ref{fig:assisted}.
Therefore, nothing changes from the previous case without
time-dependent perturbation unless the solution of the equation of
motion reaches $\omega x_4=\pi/2$.  In other words, the motion is
restricted from $u=0$ to $u^\ast$ (where $u^\ast<1$) defined by
$\omega x_4(u^\ast)=\pi/2$ above the threshold.  The critical
condition for this threshold is,
$(x_4)_{\rm max}=(m/eE)\ge \pi/2\omega$, that is,
$\gamma\equiv m\omega/(eE) \ge \pi/2$.  It is extremely interesting
that the action is modified even though the magnitude of the
time-dependent perturbation $\propto\varepsilon$ is arbitrarily
small.  Actually, the exponential factor, $\rme^{-S}$ appearing in the
pair production formula, is suppressed above the threshold
as~\cite{Schutzhold:2008pz}
\begin{equation}
  S \simeq \frac{\pi m^2}{eE}\cdot \frac{2}{\pi}\Biggl[
    \sin^{-1}\biggl(\frac{\pi}{2\gamma}
     \biggr) +\biggl(\frac{\pi}{2\gamma}\biggr)\sqrt{1-\biggl(
       \frac{\pi}{2\gamma}\biggr)} \Biggr] \qquad
  \text{for}\quad \gamma\ge\frac{\pi}{2}\,,
\end{equation}
and this reduced exponent is characteristic to what is called the
dynamically assisted Schwinger Mechanism.  For consistency check, at
the threshold $\gamma=\pi/2$, we note that the above expression
recovers the familiar Schwinger result, $S=\pi m^2/(eE)$.

\subsection{Chiral Anomaly and Axial Ward Identity}
\label{sec:anomaly}

In this subsection we will address an application of the Schwinger
Mechanism to investigate the chiral anomaly.  If the fermion mass is
zero in the Dirac Lagrangian density, the axial vector current should
be a conserved N\"{o}ther current on the classical level.  With quantum
corrections, however, the conservation law of the axial vector current
is not compatible with the gauge invariance.  Because the gauge
symmetry must not be broken (otherwise, the renormalizability is
damaged), the axial vector conservation should receive a correction
which breaks down the classical conservation law.  For given gauge
background the violation of the conservation law is precisely
quantified in a way known as the axial Ward identity:

\[
\framebox{
  \begin{minipage}{0.8\textwidth}
    Axial Ward Identity: The divergece of the axial vector current
    $j_5^\mu$ is zero if all the fermion masses are zero on the
    classical level, which is modified by quantum corrections.  If the
    theory has one Dirac fermion and Abelian gauge background fields,
    the violation is formulated on the level of the operator identity
    as
    \begin{equation}
      \partial_\mu j_5^\mu = -\frac{e^2}{16\pi^2}
      \epsilon^{\mu\nu\alpha\beta} F_{\mu\nu}F_{\alpha\beta}
      +2m\bar{\psi}\rmi \gamma_5\psi
      \label{eq:awi}
    \end{equation}
  \end{minipage}
}
\]

In the presence of parallel $\bE$ and $\bB$, therefore, the
expectation values taken on the axial Ward identity read:
\begin{equation}
  \partial_t \langle j_5^0\rangle = \frac{e^2 EB}{2\pi^2}
  + 2m\langle\bar{\psi}\rmi \gamma_5\psi\rangle\,,
\label{eq:awi_EB}
\end{equation}
where we used $\langle j_5^i\rangle = 0$ (which can be checked by
explicit calculations).  This expression has an
interesting interpretation.  Let us assume that we can drop the last
term in the chiral limit of $m\to 0$ (whose validity is far from
trivial as we will argue later).  Then, we can derive the above
relation from a very classical consideration.

\begin{figure}
  \centering
  \includegraphics[width=0.25\textwidth]{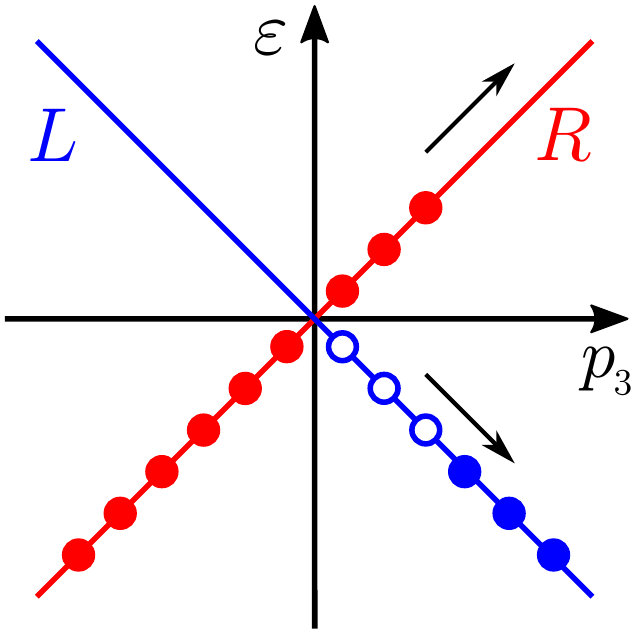}
  \caption{LLLA dispersion relations and the motion of
    particles/anti-particles in response to external electric field.}
  \label{fig:anomaly}
\end{figure}

For sufficiently strong background of $\bB$, the dimensional reduction
occurs and the (1+1)-dimensional fermionic dispersion relations belong
to the right-handed and the left-handed helicities as shown in
Fig.~\ref{fig:anomaly}, which is essentially the same figure found in
Ref.~\cite{Nielsen:1983rb}.  The Fermi momentum of the right-handed
particles increases with $E$ and its phase space volume gives the
increasing rate of the right-handed number density, $n_{\rm R}$, as
\begin{equation}
  \frac{\partial n_{\rm R}}{\partial t}\biggr|_{\text{(1+1)D}}
  = \frac{\partial}{\partial t} \frac{eE t}{2\pi}
\end{equation}
in reduced (1+1)-dimensional dynamics.  For the left-handed particles
the number density decreases, whose rate is given by the above
expression with the opposite sign.  The physical interpretation of
Fig.~\ref{fig:anomaly} is transparent;  under strong $B$ together with
$E$ the pair creation produces a right-handed particle and a
left-handed anti-particle, incrementing the chirality by two.  Then,
the total particle number is conserved, but the chirality density,
$n_5=n_{\rm R}-n_{\rm L}$, changes in (3+1) dimensions as 
\begin{equation}
  \frac{\partial n_5}{\partial t} = \frac{e^2EB}{2\pi^2}
\end{equation}
multiplied by the transverse phase space by the Landau degeneracy
factor, $eB/(2\pi)$.  This result perfectly agrees with
Eq.~\eqref{eq:awi_EB} if $n_5$ is $\langle j_5^0\rangle$.

More importantly, the above mentioned derivation of the chiral anomaly
based on the pair production of a right-handed particle and a
left-handed anti-particle leads to a quite suggestive relation, that
is,
\begin{equation}
  2w \;\;\simeq\;\; \partial_t \langle j_5^0\rangle
  = \frac{e^2 EB}{2\pi^2} + 2m\langle\bar{\psi}\rmi\gamma_5
  \psi\rangle\,,
\label{eq:conjecture}
\end{equation}
where $w$ represents the Schwinger pair production
rate~\eqref{eq:Schwinger}.

\subsubsection{In- and out-states}

If the particle mass is zero, i.e., $m=0$, it seems that
Eq.~\eqref{eq:conjecture} appears consistent with
Eq.~\eqref{eq:Schwinger}.  This statement is what is frequently said
in the literature, but we will see that the fact is far more
complicated.  To this end, we need to evaluate
$\langle\bar{\psi}\rmi\gamma_5\psi\rangle$.

It was Schwinger~\cite{Schwinger:1951nm} who first calculated
$\langle\bar{\psi}\rmi\gamma_5\psi\rangle$ using his proper-time
integration technique.  In the vacuum such a pseudo-scalar condensate
is vanishing, but with $\bE\cdot\bB$ being $\calP$-odd and
$\calC\calP$-odd, a finite expectation value should be induced as
$\propto \bE\cdot\bB$.  Then, the direct calculation results in
\begin{equation}
  \langle\bar{\psi}\rmi\gamma_5\psi\rangle =
  -\frac{e^2 EB}{4\pi^2 m}
\end{equation}
for constant and parallel electromagnetic field without
approximation.  By substituting this for the axial Ward identity, we
have to conclude that
\begin{equation}
  \partial_t \langle j_5^0\rangle = 0
  \label{eq:zero}
\end{equation}
for any $m$ including the $m\to 0$ limit.  This result is astonishing,
though this is correct!  First, Eq.~\eqref{eq:zero} obviously
contradicts an expected relation~\eqref{eq:conjecture}.  Second, the
last term $\propto m$ in Eq.~\eqref{eq:awi} can not always be dropped
in the chiral limit of $m\to 0$.  Because the condensate behaves like
$\propto m^{-1}$, the $m$-dependence in the term $\propto m$ cancels
out with the condensate and such a combination may survive.  Third,
the right-hand side of the axial Ward identity is zero, so that the
axial vector current can be conserved.  There is no way to access the
chiral anomaly which existed on the operator level but disappears as
an expectation value.

\begin{figure}
  \centering
  \includegraphics[width=0.6\textwidth]{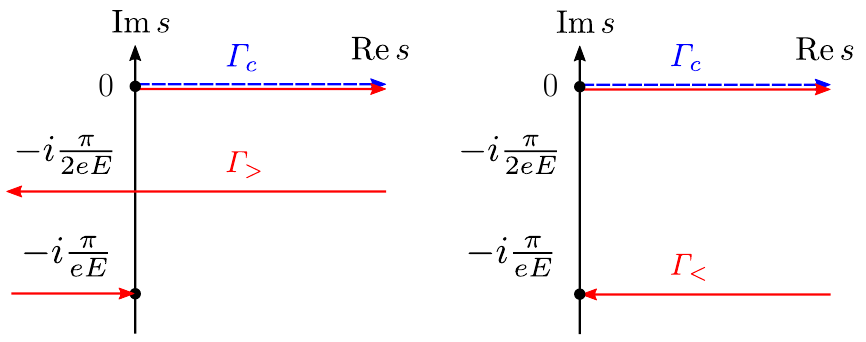}
  \caption{Contour $\Gamma_c$ for
    $\langle{\rm out}|\dots|{\rm in}\rangle$ and two deformed contours,
    $\Gamma_>$ and $\Gamma_<$, necessary for
    $\langle{\rm in}|\psi(x)\bar{\psi}(y)|{\rm in}\rangle$.
  For $x_3-y_3>0$ (and $x_3-y_3<0$), $\Gamma_>$ (and $\Gamma_<$,
  respectively) is chosen.  The presented paths may look different
  from the contours in Ref.~\cite{fradkin1991quantum} but these are
  equivalent.  Figure taken from Ref.~\cite{Copinger:2018ftr}.}
  \label{fig:contour}
\end{figure}

We can resolve this apparent puzzle once we notice that there are
inequivalent ways to take the expectation values in the presence of
electric field.  That is, as we already learnt from
Fig.~\ref{fig:E_profile}, an electric field causes a difference
between the in-state, $|{\rm in} (t\!\to\!-\infty)\rangle$, and the
out-state, $|{\rm out} (t\!\to\!+\infty)\rangle$, which makes them
distinct states.  In the standard calculus of Schwinger's proper-time
integration, the expectation value generally calculable in Euclidean
theory corresponds to $\langle{\rm out}|\cdots |{\rm in}\rangle$.
Strictly speaking, this expectation value is not a physical
observable, but an amplitude whose squared quantity is given an
interpretation as a physical observable.  To put it another way,
$\langle{\rm out}|\cdots|{\rm in}\rangle$ is just an expectation value
in Euclidean theory which is naturally realized in the $T\to 0$ limit
of finite-temperature quantum field theory.  So, we may say that
$\langle{\rm out}|\cdots|{\rm in}\rangle$ is a static or spatial
expectation value.  We thus need to cope with
$\langle{\rm in}|\cdots |{\rm in}\rangle$ in order to access the
dynamical or temporal properties of the problem.

In Ref.~\cite{Copinger:2018ftr} it has been pointed out that a
textbook~\cite{fradkin1991quantum} developed convenient technologies
for the treatments of
$\langle{\rm out}|\cdots|{\rm in}\rangle$ and
$\langle{\rm in}|\cdots|{\rm in}\rangle$.  The conventional Schwinger
proper-time integration goes on $\Gamma_c$ as depicted in
Fig.~\ref{fig:contour}, which yields
$\langle{\rm out}|\cdots|{\rm in}\rangle$.  For
$\langle{\rm in}|\psi(x)\bar{\psi}(y)|{\rm in}\rangle$ of our current
interest, we should choose the deformed contours, $\Gamma_>$ for
$x_3-y_3>0$ and $\Gamma_<$ for $x_3-y_3<0$.  Therefore, the induced
pseudo-scalar condensate has the following representation:
\begin{equation}
  \lim_{y\to x}\langle{\rm in}|\bar{\psi}(y)\rmi\gamma_5\psi(x)|{\rm in}\rangle
  =-4\rmi \frac{m e^2 EB}{(4\pi)^2}\int_{\Gamma_{\gtrless}} \rmd s\,
  \rme^{-\rmi m^2 s} = -\frac{e^2 EB}{4\pi^2 m}
  \Bigl[ 1-\rme^{-\pi m^2/(eE)} \Bigr]\,,
\label{eq:inin_ps}
\end{equation}
where $\Gamma_{\gtrless}$ is either $\Gamma_>$ or $\Gamma_<$;  because
the integrand contains no singularity, both $\Gamma_{\gtrless}$ give
a unique answer.  The above answer is consistent with the discussions
also in Ref.~\cite{Warringa:2012bq} in which
$\langle{\rm in}|\cdots|{\rm in}\rangle$ quantities have been
addressed for the field-theoretical computation of $w$, that is the
left-hand side of Eq.~\eqref{eq:conjecture}.  For the
field-theoretical formulation of $w$, see also
Ref.~\cite{Fukushima:2009er}.  Now, it is clear that
Eq.~\eqref{eq:conjecture} holds with Eq.~\eqref{eq:inin_ps} as it
should.

\subsubsection{Electromagnetic realization of the chiral magnetic effect}
\label{sec:cme}

We have seen that the Schwinger Mechanism describes a physical process
of pair production of a right-handed particle and a left-handed
anti-particle if applied $B$ is strong enough.  Apart from the
backreaction, therefore, the pair production induces a chiral
imbalance onto the system.  It is known that such a chiral imbalance
coupled with external magnetic field would be a source of exotic
phenomenon in connection to the chiral anomaly.

Before the application of the Schwinger Mechanism, we shall make a
flash overview of the chiral magnetic effect (CME), that is a
topologically induced signature for the chiral anomaly.  There are
many derivations and arguments, but one of the clearest passages
leading to the CME formula is the Maxwell-Chern-Simons theory (aka
axion electrodynamics), that is defined by the Maxwell theory augmented
with the topological $\theta$ term:
\begin{equation}
  \calL = -\frac{1}{4}F_{\mu\nu}F^{\mu\nu}
  -\frac{e^2}{16\pi^2}\theta\, F_{\mu\nu}\tilde{F}^{\mu\nu}
  +\bar{\psi}(\rmi\feyn{D}-m)\psi\,.
\label{eq:CSM}
\end{equation}
The second term involving the $\theta$ angle is the Chern-Simons term,
and as long as $\theta$ is constant, this $\theta$ term would not
modify the equation of motion because $F_{\mu\nu}\tilde{F}^{\mu\nu}$
is a total derivative.  If we assume spacetime dependent $\theta$,
however, some additional terms appear in the equation of motion and
the modified Gauss and Amp\`{e}re laws read~\cite{Wilczek:1987mv}:
\begin{align}
  &\text{(Gauss law)} \qquad
  \bnabla\cdot\bE = \rho + \frac{e^2}{4\pi^2}
  (\bnabla\theta)\cdot\bB\,,
  \label{eq:gauss}\\
  &\text{(Amp\`{e}re law)} \qquad
  \bnabla\times\bB - \dot{\bE} = \bj + \frac{e^2}{4\pi^2}
  \bigl[ \dot{\theta}\bB - (\bnabla\theta)\times\bE \bigr]\,.
  \label{eq:ampere}
\end{align}
From this expression, we see that new terms arise in the place of
$\bj$ which can be identified as new contributions to the current.
One might think that such induced terms could be a fictitious current
like Maxwell's displacement current, $\dot{\bE}$, as questioned in
Ref.~\cite{Fukushima:2012vr}.  To answer this question, we should
consider how the charge conservation holds for a finite system; actually,
charge density $\propto (\bnabla\theta)\cdot\bB$ emerges in a
perfectly consistent way with the current $\propto \dot{\theta}\bB$,
which means that this new current $\propto \dot{\theta}\bB$ is a
genuine current unlike Maxwell's displacement current.  Changing the
notation from $\dot{\theta}$ to $\mu_5$, we arrive at the standard CME
formula~\cite{Fukushima:2008xe} (for a pretty complete list of
preceding works including Vilenkin~\cite{Vilenkin:1980fu} and
Giovannini-Shaposhnikov~\cite{Giovannini:1997gp},
see Ref.~\cite{Kharzeev:2015znc}):

\[
\framebox{
  \begin{minipage}{0.8\textwidth}
    Chiral Magnetic Effect: A nonzero chirality imbalance is induced
    by $\dot{\theta}=\mu_5$, which is coupled with external magnetic
    field, $\bB$, leading to an electric current parallel to $\bB$ as
    \begin{equation}
      \bj_{\rm CME} = \frac{e^2}{4\pi^2}\mu_5 \bB
    \end{equation}
  \end{minipage}
}
\]

This is a formula for a single fermion contribution.  If the theory
has multiple species of fermions, the above current is multiplied by
the fermionic degrees of freedom.  Interestingly, the formula is
insensitive to the fermion mass, and it is independent of the
temperature, as should be so for the quantum anomaly.

\begin{figure}
  \centering
  \includegraphics[width=0.5\textwidth]{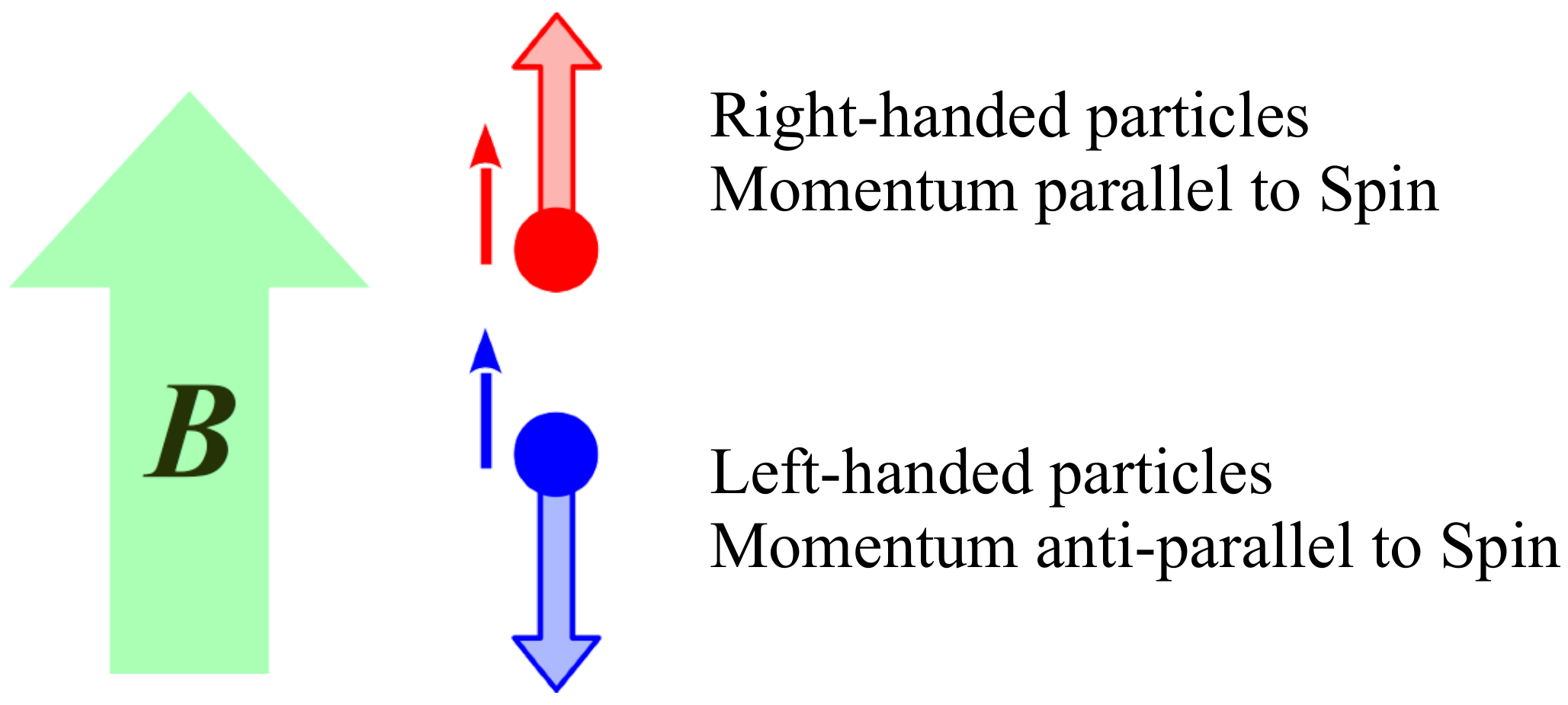}
  \caption{A classical picture for the CME under a strong magnetic
    field.  The spin is aligned to the magnetic direction, which
    uniquely fixes the direction of the momentum depending on whether
    particles are right-handed or left-handed.  Then, if there is an
    excess of either right-handed or left-handed particles, a net
    electric current flows along the applied magnetic field.}
  \label{fig:cme}
\end{figure}

An intuitive view point to understand the CME is picturized in
Fig.~\ref{fig:cme}.  By definition, the right-handed
particles have the spin and the momentum in parallel directions, while
the left-handed particles have anti-parallel spin and momentum.  The
spin direction is aligned along the magnetic field as shown in
Fig.~\ref{fig:cme}, which simultaneously fixes the momentum
direction.  A nonzero $\mu_5$ induces unbalanced chirality density,
$n_5=n_R-n_L\neq 0$, and thus, a combination of $n_5\neq 0$ and the
spin-magnetic correlation generates a net current along the magnetic
direction.  Such an electric current $\parallel \bB$ is inherently
anomalous that does not exist in classical electrodynamics in which
the magnetic field does not transfer any work on charged particles.
The energy is supplied by $\mu_5$ or in other words the driving force
is provided by time-dependent $\theta$.

Regarding the physical interpretation of the formula, there have been
controversies.  Some came from the derivation presented in
Ref.~\cite{Fukushima:2008xe};  the thermodynamic potential
$\Omega[\bA]$ was calculated on background vector potential $\bA$,
magnetic field $\bB$, and chiral chemical potential $\mu_5$, and the
current was estimated from $\bj\propto\delta\Omega/\delta\bA$.
Nothing is wrong about this procedure, but strangely, this derivation
implies the existence of the permanent current even in equilibrium!  A
state with current should be, even if it is time independent, a steady
state out of equilibrium.  In short, in equilibrium, the CME current
must be prohibited by definition of equilibrium.  Here, one must not
be confused with the argument in Ref.~\cite{PhysRevLett.111.027201},
though the claim itself sounds similar.  The important message from
Ref.~\cite{PhysRevLett.111.027201} is that the CME current does not
exist in solid state systems where there is no Dirac sea (from which
the anomaly arises) and also the surface term cancels out with
periodic boundary condition (see also Ref.~\cite{Yamamoto:2015fxa} for
related discussions).  In this sense, the statement of
Ref.~\cite{PhysRevLett.111.027201} is the same claim but made under
tighter constraints, not applied to continuous quantum field theory.
What I am emphasizing here is that even in continuous quantum field
theory the CME current cannot be permitted in equilibrium.

A sort of consensus has been built in the community, that is, a
nonzero $\mu_5$ is the trick to emulate a steady state using an
equilibrium language.  More concretely, $\mu_5$ should be zero in true
equilibrium situation, and the introduction of nonzero $\mu_5$ forces
the system to be out of equilibrium.  Such an interpretation
physically makes sense, but as long as $\mu_5$ is introduced by hand,
there is no way to turn off effects to mimic off-equilibrium.  One way to
cure this situation is to take account of the relaxation process of
$\mu_5$ decaying toward zero as the time goes.

Another (and more well-founded) theoretical approach would be to
abandon using $\mu_5$.  The role played by $\mu_5$ is to inject an
energy imbalance between the right-handed particle sector and the
left-handed particle sector.  We have already argued that the
Schwinger Mechanism under strong magnetic field leads to exactly such
chirality imbalance.  Hence, we can just impose an external $\bE$ so
that $\bE\cdot\bB\neq 0$ can replace $\mu_5$.  This idea was first
proposed in Ref.~\cite{Fukushima:2010vw} and analytical formulas based
on the identification of Eq.~\eqref{eq:conjecture} have been derived
for a geometry of $\bE$ and $\bB$ making an angle less than $\pi/2$
(i.e., $\bB$ has both components $\parallel\bE$ and $\perp\bE$).  We
have already seen that, surprisingly, the Euclidean expectation values
make the right-hand side of the axial Ward identity vanishing, which
also means that the CME current is vanishing then.  Now, we have a
better understanding of why this should be so;  the CME current should
be indeed vanishing for static systems described in Euclidean
formulation, as was clarified recently in Ref.~\cite{Copinger:2018ftr}.

\begin{figure}
  \centering
  \includegraphics[width=0.35\textwidth]{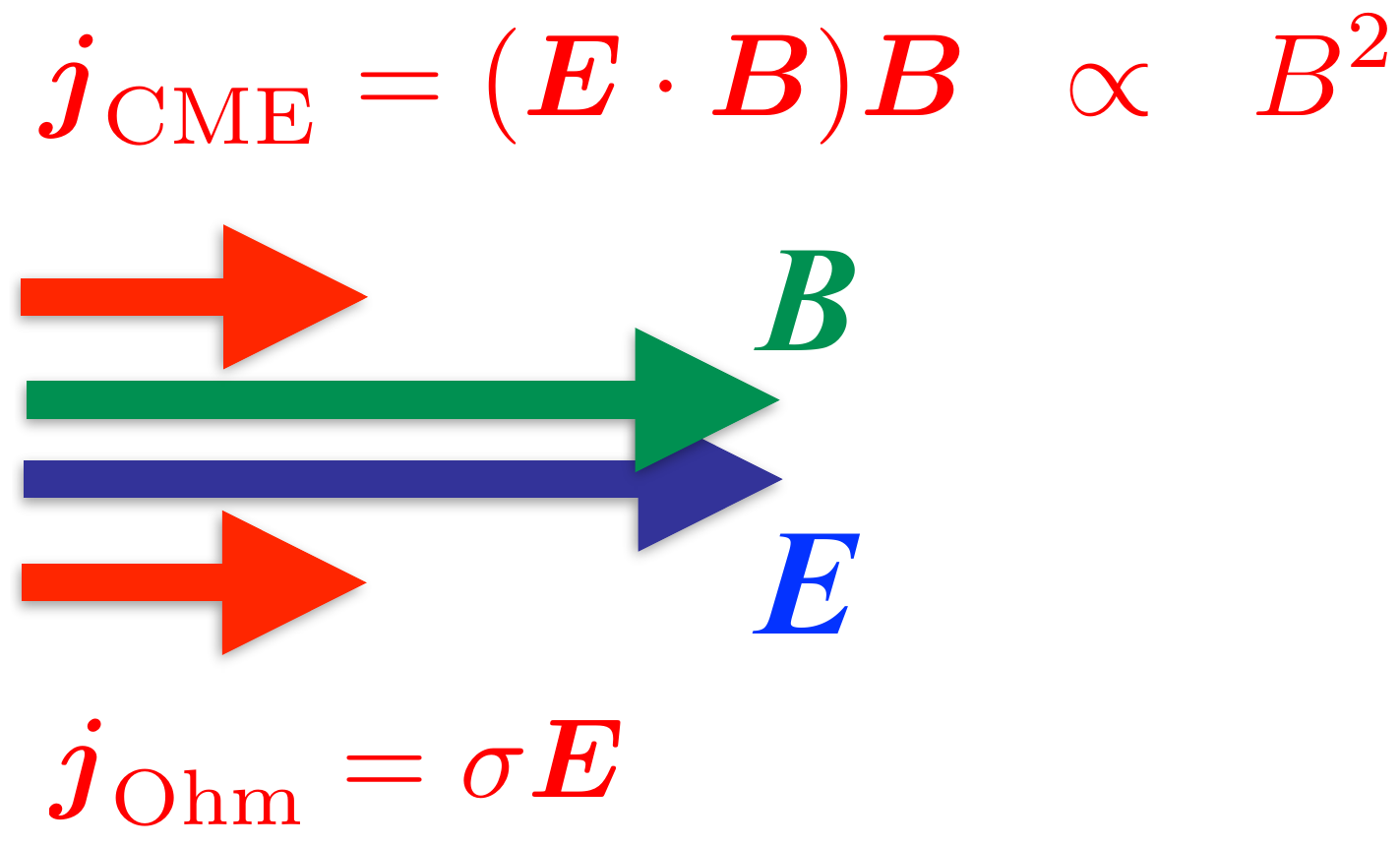}
  \caption{Simplest setup to observe the CME current in terms of the
    negative magnetoresistance.}
  \label{fig:negative_mr}
\end{figure}

Later on, the simple idea of Ref.~\cite{Fukushima:2010vw} was further
simplified in Ref.~\cite{Son:2012bg};  no angle between $\bE$ and
$\bB$ is necessary, but only the parallel $\bE\parallel\bB$ is
sufficient to probe the CME current as sketched in
Fig.~\ref{fig:negative_mr}.  The idea is the following.  The minimal
setup is just to impose an external magnetic field on chiral material
along which a finite voltage is applied so that an electric current
flows.  Then, the electric field should be present there according to Ohm's law, i.e.,
$\bj_{\rm Ohm}=\sigma\bE$ where $\sigma$ denotes the Ohmic electric
conductivity.  Because of the electromagnetic background, a finite
chirality is pumped up as $n_5\propto \bE\cdot\bB$ as deduced from
Eq.~\eqref{eq:Schwinger}, which means that $\mu_5\propto \bE\cdot\bB$
is anticipated.  Then, the CME formula with
$\mu_5\propto \bE\cdot\bB$ implies the CME current,
$\bj_{\rm CME}\propto (\bE\cdot\bB)\bB\propto B^2$.  From this relation we
can deduce that the electric conductivity $\sigma_{\rm CME}$
associated with the CME current has peculiar magnetic dependence
$\propto B^2$.  The experimentally observed current should be a
superposition of $\bj_{\rm Ohm}+\bj_{\rm CME}=(\sigma+\sigma_{\rm CME})\bE$,
where $\sigma$ is assumed to have only minor dependence on $\bB$.
Therefore, if an anomalous component of the electric conductivity
increasing with $B^2$ or equivalently the electric resistivity
decreasing with $B^{-2}$ is confirmed, it would be a clear
experimental signature for $\sigma_{\rm CME}$ and thus the chiral anomaly.  Such a behavior of the
resistivity suppressed as $B^{-2}$ is especially referred to as the
``negative magnetoresistance'' in the literature.  Finally, let us
make a remark on the asymptotic $B$ dependence in the strong-$B$
limit.  In the above discussion a relaxation process of the produced
chirality is implicitly assumed;  otherwise, the chirality diverges
over a long time.  Actually in Ref.~\cite{Son:2012bg} the relaxation time
approximation is adopted in the framework of the chiral kinetic
theory, and also in Ref.~\cite{Li:2014bha} that is the very first report of
the experimental observation of the negative magnetoresistance, a
relaxation time is introduced in a Drude-type picture.  Then, such
quadratic dependence of the conductivity or resistivity is predicted
if the relaxation time is $B$ independent.  In reality, however, the
relaxation time or the microscopic scattering process is
significantly affected by $B$-dependent phase space volume.  A
complete field-theoretical computation of the electric conductivity
including higher Landau levels have revealed that the physical
asymptotic dependence on $B$ is not quadratic but linear~\cite{Fukushima:2017lvb}.

\section{Phenomena induced by Rotation}
\label{sec:rotation}

In many physics problems the angular momentum causes a
phenomenological consequence analogous to the magnetic effect.
In fact rotating chiral matter would exhibit a current which looks
similar to the CME current.  This effect to generate rotation induced
axial current is called the chiral vortical effect (CVE).  Because the
axial current operator is nothing but the spin operator in the
relativistic language, the CVE could be regarded as a transport
process from the mechanical rotation into the spin or the
magnetization, which may well be a relativistic extension of the
Barnett effect.  Then, it would be a natural question to think of
another relativistic extension, namely, the Einstein--de-Haas effect.  These
are new subjects and so we will only briefly look over speculative
ideas.

Furthermore, recently, there are many interesting works to discuss the
ground state properties affected by rotation effects, see
Ref.~\cite{Jiang:2016wvv} for a phase diagram,
Ref.~\cite{Chernodub:2017ref} for the scalar condensate,
Ref.~\cite{Liu:2017spl} for the pion condensation, etc.  These are all
exciting developments, but in this review, we will stay with simpler
physics problems only.

\subsection{Rotating Chiral Fermions}
\label{sec:rotating}

Rotation effect has been well understood in some fields of physics
such as nuclear physics in which all deformed nuclei must rotate to
restore broken symmetry in a finite system.  Such a quantum system in
a rotating frame can be described by the cranking Hamiltonian, that
is, the Hamiltonian shifted by a cranking term, $\bj\cdot\bomega$, where
$\bomega$ is the angular velocity vector and $\bj$ is the total
angular momentum.  Such an energy shift can be identified as an
effective chemical potential.  In this sense, a finite rotation has
dual aspects as a magnetic counterpart and a finite-density
counterpart.  In particular the analogy between rotation and finite
density has been pursued in Ref.~\cite{Chen:2015hfc} in which the
rotation-induced inverse magnetic catalysis has been found.  From this
analogy the phase diagram as a function of the temperature vs.\ the
rotation angular velocity can be considered, and similarity to the
conventional phase diagram as a function of the temperature vs.\ the
chemical potential has been verified in Ref.~\cite{Jiang:2016wvv},
with an importance exception.  Suppose that the system is just rotated
but no magnetic field nor temperature is introduced, then, the
question is as follows;  can a finite density be induced by the
effective chemical potential by $\bj\cdot\bomega$ alone?  The answer
is negative.  The finite size effect is crucial to make a correct
statement.  With discrete momenta, Eq.~\eqref{eq:disc_p}, the minimum
energy gap is of order $1/R$, and the effective chemical potential
$\bj\cdot\bomega$ is also of order $1/R$ bounded by the causality, and
one can prove that the former is always greater than the
latter~\cite{Ebihara:2016fwa}.  Therefore, only a finite rotation
cannot change the ground state property, which is consistent with the
equivalence principle.

\begin{figure}
  \centering
  \includegraphics[width=0.4\textwidth]{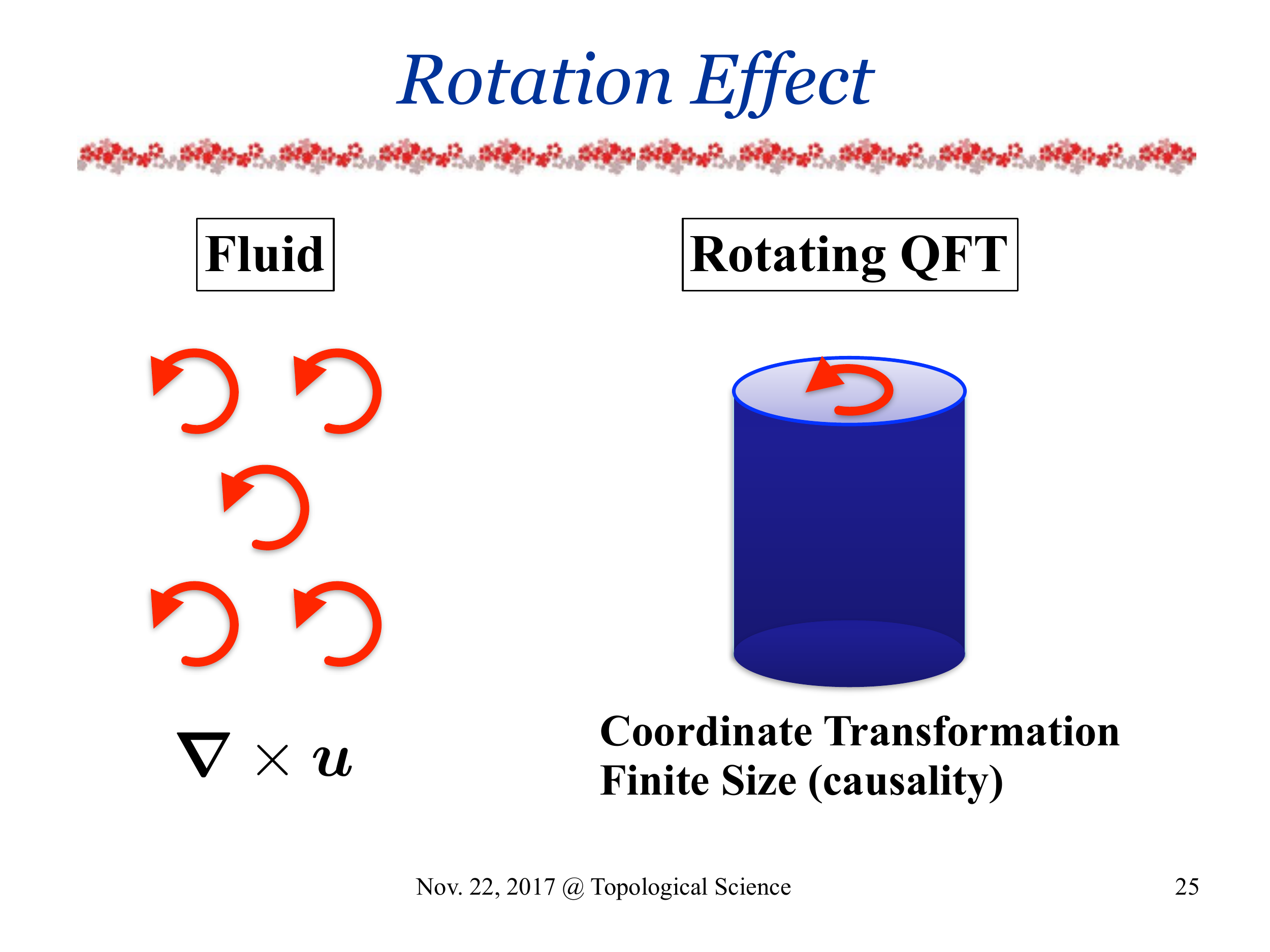}
  \caption{Two treatments of rotational effects;  (left) local
    vorticity vector characterizes the distribution of
    $\bnabla\times\boldu$ where $\boldu$ is a fluid velocity vector
    and (right) rigidly rotating object treated in a rotating frame
    gives global rotation but needs boundary not to break causality.}
  \label{fig:rotation}
\end{figure}

To incorporate rotational effect, two technically distinct treatments
are commonly adopted;  one method is a description with local
vorticity vector of a fluid, and the other is with global rotation via
coordinate transformation to a rotating frame, as schematically
drawn in Fig.~\ref{fig:rotation}.  Each has an advantage and a
disadvantage.  Because the vorticity is a local quantity, the former
method has no necessity to be put in a finite sized system.  Without
the boundary, however, there is no simple way to take account of the
orbital angular momentum (see Refs.~\cite{Becattini:2009wh} for
examples of theoretical attempts for spinning hydrodynamics).  The
latter is in this sense more convenient in order to consider the
orbital angular momentum of a finite system.  Unlike the former fluid
description, however, the center of rotation and the boundary at the
surface make physical quantities inhomogeneous depending on the radial
distance.

Here, I would stress that two treatments should be equivalent if the
same physical system is considered.  It is well known that, for
example, a rotating superfluid forms a lattice of vortices which carry
a quantized angular momentum, and as the rotating velocity of the
superfluid increases, the number of vortices gets larger, and  the
integrated vorticity eventually approaches the rigid rotor limit.  In
this way, a uniform distribution of vorticity vector should amount to
the global rotation.

\subsubsection{Field-theoretical treatments}

I prefer the treatment using quantum field theory in a rotating
frame, which enables us to perform rather brute-force calculations.  The derivation of the rotation induced current is quite
suggestive, having similarity to the Casimir effect to some extent,
which is worth discussing.

The starting point of field-theoretical computation is the Dirac
equation in a rotating frame.  For free Dirac fermion the Lagrangian
density reads:
\begin{equation}
  \bigl[ \rmi\gamma^\mu(\partial_\mu + \Gamma_\mu) - m\bigr]\psi
  = 0\,,
\end{equation}
where $\Gamma_\mu=-\frac{\rmi}{4}\omega_{\mu ij}\sigma^{ij}$ with
$\sigma^{ij}=\frac{\rmi}{2}[\gamma^i,\gamma^j]$.  The spin connection
is $\omega_{\mu ij}=g_{\alpha\beta}\,e_i^\alpha (\partial_\mu
e_j^\beta + \Gamma_{\mu\nu}^\beta\,e_j^\nu)$ given in terms of the
metric and the Vierbein.  In a frame rotating along the $z$-axis, the
metric involves the angular velocity $\omega$ as
\begin{equation}
  g_{\mu\nu} = \begin{pmatrix}
    1-(x^2+y^2)\omega^2 & y\omega & -x\omega & 0\\
    y\omega & -1 & 0 & 0 \\
    -x\omega & 0 & -1 & 0 \\
    0 & 0 & 0 & -1
  \end{pmatrix}\,.
\end{equation}
The corresponding Vierbein is not uniquely fixed but can be chosen as
\begin{equation}
  e_0^t = e_1^x = e_2^y = e_3^z = 1\,,\qquad
  e_0^x = y\omega\,,\qquad e_0^y = -x\omega\,.
\end{equation}
With this choice we can explicitly write down the free Dirac equation
in the following way:
\begin{equation}
  \biggl\{ \rmi\gamma^0 \Bigl[\partial_t+\omega\Bigl(-x\partial_y+y\partial_x
    -\frac{\rmi}{2}\sigma^{12}\Bigr)\Bigr] - \rmi\gamma^1\partial_x
  -\rmi\gamma^2\partial_y-\rmi\partial^3\partial_z - m \biggr\}\psi = 0\,.
\label{eq:Dirac_rot}
\end{equation}
Interestingly, though the metric contains $\omega^2$ term, the
Vierbein is linear and this above equation is also linear in terms of
$\omega$.  The physical interpretation of terms coupled with $\omega$
is evident.  As explained above, the cranking term shifts the
Hamiltonian or the energy by $\bj\cdot\bomega$, and the quantity with
parentheses is nothing but the total angular momentum.  This energy
shift is essential to understand various physics phenomena, like a
critical velocity for the superfluid vortex.

\subsubsection{Chiral vortical effect}

It was Vilenkin who found that the expectation value of the axial
vector current is something finite proportional to
$\omega$~\cite{Vilenkin:1978hb}, which was reconfirmed by more
field-theoretical frameworks
later~\cite{Vilenkin:1979ui,Vilenkin:1980zv}.  The statement itself is
quite nontrivial, and moreover, the derivation is extremely
interesting.  Before looking at the derivation, let us see the final
expression:

\[
\framebox{
  \begin{minipage}{0.8\textwidth}
    Chiral Vortical Effect: Axial vector current is induced by a
    combination of rotation with the angular velocity $\omega$ and
    chiral matter characterized by fermionic thermal distribution
    function $f_{\rm F}(\varepsilon_p)$.  For the distribution
    functions with chemical potentials $\mu_{R/L}$ for the
    right-handed and the left-handed fermions, the CVE formula reads:
\begin{equation}
  \bj_{R/L} = \mp \bomega \int\frac{\rmd^3 p}{(2\pi)^3}\,
  f_{\rm F}'(\varepsilon_p;\mu_{R/L})\,,
  \label{eq:cve_f}
\end{equation}
where $f_{\rm F}'(\varepsilon_p)$ represents the derivative of
$f_{\rm F}(\varepsilon)$ with respect to $\varepsilon_p$.
  \end{minipage}
}
\]
The structure of the formula may look similar to the CME current;  the
CVE current is proportional to $\bomega$ and the CME current to $\bB$,
which are both not explainable in classical physics.  However, there
is an essential difference.  The CME appears from the vacuum
fluctuation and there is no contribution from the finite-$T$ part, but
the CVE originates from the matter part only.  The above formula is
further simplified for $m=0$ as
\begin{equation}
  \bj_{R/L} = \pm\biggl(\frac{T^2}{12} + \frac{\mu_{R/L}^2}
     {4\pi^2} \biggr)\bomega
     \label{eq:cve}
\end{equation}
which is perhaps the most well recognized expression in the
literature.

Now we shall take a closer look at the derivation.  As seen from the
Dirac equation~\eqref{eq:Dirac_rot}, the rotation effect lies in an
energy shift by $\bj\cdot\bomega$, and so the propagator $S$ in a
rotating frame is obtained from $S_0$ in the inertial frame with the
shift operator as
\begin{equation}
  S(\bx,\bx',p_0) = \rme^{\bomega\cdot\bj\frac{\partial}
    {\partial p_0}}\, S_0(\bx,\bx',p_0)\,,
\end{equation}
where $\bj = \bL + \frac{1}{2}\bSigma$ with the spin vector defined as
$\Sigma^i \equiv \frac{1}{2}\epsilon^{ijk}\sigma^{jk}$.  It should be
noted that the above propagator is Fourier transformed only for the
temporal component, for $\bL$ is $\bx$ dependent.  We are now
estimating the axial vector expectation value,
$\-\rmi\tr[\gamma^\mu\gamma_5 S(x,x)]$, which involves $\gamma_5$.
This trace is vanishing unless three $\gamma$ matrices come along from
$S$.  The free propagator, $S_0$, is proportional to one $\gamma$
matrix, and two more are necessary from the shift operator,
$\rme^{\bomega\cdot\bj\frac{\partial}{\partial p_0}}$.  For small
$\bomega$ we can expand the shift operator in terms of $\bomega$ and
then the first remaining term is thus the spin part with $\bSigma$
which contains two $\gamma$ matrices.  Therefore, the expectation
value of the axial vector current reads:
\begin{equation}
  j_5^\mu = -\rmi \tr \bigl[\gamma^\mu \gamma_5 S(x,x) \bigr]
  = -\rmi \bomega\, \tr\biggl[\gamma^\mu\gamma_5
    \frac{1}{2}\bSigma \frac{\partial}{\partial p_0}\,
    S_0(\bp,p_0) \biggr] + O(\omega^2)\,.
\end{equation}
Here, the first $\tr$ is taken in configuration space and the second
in momentum space.  Using the trace property,
$\tr[\gamma_5\gamma^\mu\gamma^\nu\gamma^\rho\gamma^\sigma]
=4\rmi\epsilon^{\mu\nu\rho\sigma}$, and introducing the angular
velocity tensor, $\omega^{ij}=\frac{1}{2}\epsilon^{ijk}\omega_k$ (in
fact, the angular velocity is originally a two-rank tensor like the
angular momentum), we can simplify the above expression as
\begin{equation}
  j_5^\mu = -\rmi \epsilon^{\mu\alpha\beta\nu} \omega_{\alpha\beta}
  \int^T\frac{\rmd^4 p}{(2\pi)^4} \frac{\partial}{\partial p_4}
  \frac{p_\nu}{p^2 + m^2}\,.
\end{equation}
Here, a finite-$T$ field theory is assumed, and $p_0$ is Wick rotated
to the Matsubara frequency, $p_4=2\pi n T$.  This expression is quite
remarkable.  If $p_4$ were a continuous variable (at $T=0$) and if
there were no surface contribution (at $\mu=0$), the integral is
trivially zero.  In other words, the term $\propto\mu^2$ in
Eq.~\eqref{eq:cve} emerges from the edges of the momentum integration,
which is a common typical feature of quantum anomaly.  The term
$\propto T^2$, in contrast, seems to have a rather different origin;  this
is nonzero because of a finite discrepancy between a continuous
integral and a discrete sum, which is more reminiscent of the Casimir
energy!

As a matter of fact, it is still under theoretical disputes whether
the $T^2$-term in the CVE formula is related to quantum anomaly or
not.  From the group theoretical structure, in
Ref.~\cite{Landsteiner:2011cp}, a connection to the mixed
gravitational chiral anomaly was conjectured, which was followed by
holographic
studies~\cite{Landsteiner:2011iq,Jensen:2012kj,Jensen:2013kka}.
However, the coefficient of the $T^2$-term is not anomaly protected,
while the $\mu^2$-term is not renormalized by the
interaction~\cite{Golkar:2012kb}.  Besides, it has been clarified that
the CVE coefficient is fixed by the mixed global anomaly, not the
perturbative anomaly~\cite{Golkar:2015oxw}, but this itself is not so
surprising because the $T^2$-term is a pure finite-$T$ effect which is
in any case traced back to the compactification.  For discussions from
the point of view of hydrodynamics, see
Refs.~\cite{Son:2009tf,Buzzegoli:2017cqy}.  A more hint for a possible
relation to the gravitational anomaly may well be available from the
calculation including spatial curvature as well as rotation.  It is a straightforward
generalization to include the effect of curved space into the above
calculation, which after some algebra leads to~\cite{Flachi:2017vlp}
\begin{equation}
  \bj_{R/L} = \pm \biggl(\frac{T^2}{12}-\frac{m^2}{8\pi^2}
    -\frac{R}{96\pi^2}\biggr)\bomega\,,
\end{equation}
apart from the chemical potential terms, together with a mass
correction up to $m^2$ order.  The last term is purely geometrical and
its coefficient not surprisingly coincides with that of the
Chern-Simons current associated with the gravitational anomaly.  The
second term is a mass correction from the distribution function in
Eq.~\eqref{eq:cve_f}.  Suggestively, this $m^2$-term from the CVE
formula is perfectly consistent with what is expected from the chiral
gap effect, that is, a mass shift by the curvature as
$m^2\to m^2+R/12$ as found in Ref.~\cite{Flachi:2014jra}.  In this
way, through the $m^2$-term, we could observe at least an indirect
relationship between the $T^2$-term in the CVE current and the
gravitational anomaly.

\subsection{Floquet Theory}
\label{sec:floquet}

The CVE is as interesting as the CME, but controlling physical rotation
in laboratory would require a delicate design for experimental machineries.
Here, we will see that, not the CVE itself, but some quite similar
phenomena would be accessible by means of circularly polarized
electromagnetic fields.  In fact, such an optical setup has been
intensely investigated in quantum optics and laser physics (and also
in the quantum field theory context too, see
Ref.~\cite{Chernodub:2018era} for a recent work), and it has been
known that the Floquet theory is a powerful tool (for a pedagogical
review on the Floquet theory, see Ref.~\cite{Hanggi:1997}).

The Floquet theory is a temporal version of the Bloch theorem, that is:
\[
\framebox{
  \begin{minipage}{0.8\textwidth}
    Floquet Theory: Quantum states for a time periodic Hamiltonian
    with a periodicity $T$, satisfying
    $H(t+T)=H(t)$, are described by wave-functions in a form of
    \begin{equation}
      \psi_n(t) = u_n(t)\,\rme^{-\rmi \varepsilon_n t}\,,
    \end{equation}
    with periodic Floquet modes, $u_n(t+T)=u_n(t)$.
  \end{minipage}
}
\]
For the time-periodic system the one-cycle time evolution defines an
effective Hamiltonian, called the Floquet Hamiltonian, given by
\begin{equation}
  \rme^{-\rmi T H_{\rm eff}}
  = \calT \exp\biggl[-\rmi\int_0^T \rmd t\,H(t)\biggr]\,.
\end{equation}
With this effective Hamiltonian the dynamical problem is reduced to a
static one.  The idea of the Floquet engineering is to design
$H_{\rm eff}$ that has desired properties on a finite time scale.  For
example spacetime dependent $\theta$ as in Eq.~\eqref{eq:CSM} could be
engineered, which would open an opportunity to study induced terms in
Eqs.~\eqref{eq:gauss} and \eqref{eq:ampere} experimentally.

\subsubsection{Rotating frame and Magnus expansion}

Here is one example of the Floquet engineering to induce $z$ dependent
$\theta$, which is equivalent to an axial vector potential
$\bA_5=\beta\ez$.  In electromagnetism, of course, there is no such
axial vector potential, and we will see a theoretical benefit from the
Floquet engineering which is clear in the demonstration presented below.

\begin{figure}
  \centering
  \includegraphics[width=0.3\textwidth]{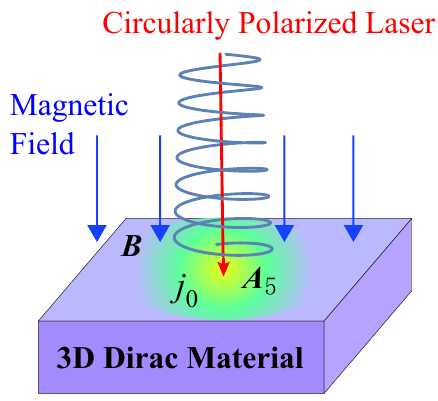}
  \caption{Chiral pumping effect:  chiral matter exposed to circularly
    polarized transverse electric field together with longitudinal
    magnetic field.  Figure taken from Ref.~\cite{Ebihara:2015aca}.}
  \label{fig:chiral_pump}
\end{figure}

Let us consider a circularly polarized electric field, as depicted in
Fig.~\ref{fig:chiral_pump}, with the following vector potential:
\begin{equation}
  A_x = \frac{E}{\omega}\cos(\omega t)\,,\quad
  A_y = \frac{E}{\omega}\sin(\omega t)\,.
\end{equation}
The corresponding Hamiltonian is periodically time-dependent, i.e.,
\begin{equation}
  H = \gamma^0\bgamma\cdot\bp + \gamma^0 m
    - e\gamma^0\bgamma\cdot\bA
    = \underbrace{\gamma^0\bgamma\cdot\bp + \gamma^0 m}_{H_0} \;\;
    \underbrace{- \frac{eE}{\omega}
      \gamma^0\gamma^- \rme^{\rmi \omega t}}_{H_-} \;\;
    \underbrace{- \frac{eE}{\omega}
  \gamma^0\gamma^+ \rme^{-\rmi \omega t}}_{H_+}
\end{equation}
with $\gamma^\pm\equiv \frac{1}{2}(\gamma^1\pm\rmi\gamma^2)$.  In the
above the whole Hamiltonian is split into a static part, $H_0$, and
the positive- and negative-frequency parts, $H_\pm$.  The Floquet
effective Hamiltonian can be systematically obtained by the
$\omega^{-1}$ expansion if $\omega$ is large enough (only for which
the Floquet theory physically makes sense; otherwise, the temperature
diverges without prethermalization~\cite{PhysRevLett.116.120401}).
Then, there is a formula known as the Magnus
expansion~\cite{BLANES2009151}, from which we can read the first
expanded term up to $O(\omega^{-1})$ using the above notation, $H_0$
and $H_\pm$, as
\begin{equation}
  H_{\rm eff} \simeq H_0 + \frac{1}{\omega}[H_-,H_+]
  = \gamma^0\bgamma\cdot\bp + \gamma^0 m - \beta\gamma^0 \gamma^z
  \gamma_5
\end{equation}
with $\beta\equiv (eE)^2/\omega^3$.  Now we understand that we can
regard this last term as an induced axial vector potential,
$\bA_5 = \beta\ez$.

Because $2\bA_5$ is transformed into $\bnabla\theta$, the modified Gauss
law~\eqref{eq:gauss} indicates that a finite charge density is pumped
up (i.e., chiral pumping effect) as
\begin{equation}
  \rho_{\rm induced} = \frac{e}{2\pi^2}\beta B\,,
\end{equation}
coupled with a magnetic field, $\bB=B\ez$, which can be derived also
from direct calculations~\cite{Ebihara:2015aca}.

The effect of circularly polarized electric field mimics rotational
effect, which can be better understood in the following manner.  Instead of
a combination of circularly polarized electric field and static
magnetic field, let us consider just a rotating magnetic field which
produces periodically driven potential.  We can eliminate time
dependence by the coordinate transformation:
\begin{equation}
  x \;\to\; (\cos\omega t)x - (\sin\omega t)y\,,\quad
  y \;\to\; (\cos\omega t)y + (\sin\omega t)x\,.
\end{equation}
Then, the Floquet transformation of the basis,
\begin{equation}
  \psi\;\to\; \exp(\gamma^1\gamma^2\omega t/2) \psi\,,
\end{equation}
with the replacement of $\gamma^i\to\gamma^\mu$ with the Vierbein
reduces the problem into a static one characterized by the transformed
Lagrangian density:
\begin{equation}
  \calL = \bar{\psi}\Bigl[ \rmi\gamma^0\partial_t
    +\rmi\gamma^1\partial_x + \rmi\gamma^2(\partial_y
    - \rmi eB x) + \rmi\gamma^3\partial_z + \frac{\omega}{2}
    \gamma^3\gamma_5 \Bigr]\psi\,.
\end{equation}
Now, we see that a rotating magnetic field is transformed into a
static one, as expected from the frame change, and the effect is not
only that but also the transformation gives rise to a new term, which
can be identified with the axial vector potential corresponding to
$\beta=\omega/2$, leading to~\cite{Ebihara:2016fwa}
\begin{equation}
  \rho_{\rm induced} = \frac{eB\omega}{4\pi^2}\,.
\end{equation}
In this case the above result assumes no approximation, which makes a
sharp contrast to the previous calculation based on the Magnus
expansion.  This exactness of resulting $\rho_{\rm induced}$ is
explained by the anomaly nature as argued in
Ref.~\cite{Hattori:2016njk} where the same expression was obtained.

\subsubsection{Interpretation as an artificial electric field}

The Floquet theory has an interesting interpretation in physics, as
slightly mentioned in a review~\cite{oka}, which
has much to do with the main subject of this article.  We will briefly
look over the basic idea of what is called Sambe
space~\cite{PhysRevA.7.2203}.

The periodic Hamiltonian has a Fourier expanded form,
\begin{equation}
  H(t) = H(t+T) = \sum_{n=-\infty}^\infty H_n\,\rme^{-\rmi n\omega t}
\end{equation}
with $\omega=2\pi/T$.  The Floquet state having a quasi-energy,
$\varepsilon_\alpha$, can be also Fourier expanded as
\begin{equation}
  u_\alpha(t) = \sum_{n=-\infty}^\infty u_{\alpha n}\, \rme^{-\rmi n\omega t}\,.
\end{equation}
Then, the eigenvalue equation to determine the quasi-eigenenergy and
the Floquet basis states reads:
\begin{equation}
  \sum_m \bigl( H_{n-m} - n\omega\delta_{mn} \bigr)u_{\alpha m}
  = \varepsilon_\alpha u_{\alpha n}\,.
\end{equation}
In Floquet picture this eigenvalue equation is given an interesting
interpretation.  The matrix representation of the quasi-energy
operator in the left-hand side has diagonal components, $H_0$,
$H_0\pm \omega$, $H_0\pm 2\omega, \dots$, and these states are
communicated by off-diagonal interactions, $H_{\pm 1}$,
$H_{\pm 2}, \dots$.  Then, along the Floquet direction labeled by
$m$, the diagonal Hamiltonian has an energy slope by $m\omega$.  If we
regard this Floquet direction as an extra spatial axis, we can
identify the energy shift, $m\omega$, as a coordinate dependent static
potential.  Therefore, the slope of the energy shift is nothing but a
static electric field $\propto \omega$ (one more mass dimension should
be compensated by ``lattice spacing'' which is arbitrary in the
present description).  This picture of fictitious electric field is
useful, for example, to think of topological aspects in Floquet
systems.

It would be a fascinating possibility to emulate effects equivalent
to $\bE\cdot\bB\neq 0$ in (3+1) dimensions without using $\bE$ but
with just time periodic perturbations onto a system in (2+1)
dimensions. Such a connection between the Floquet system and the
finite-$\bE$ system should deserve more investigations from both
theoretical and experimental sides.

\subsection{Relativistic Gyromagnetic Effects}

The CVE is an axial vector current induced by rotation, as we have
already seen, and the axial vector is translated to a spin expectation
value in the non-relativistic limit.  Actually, in non-relativistic
theories, it is a well established notion that a finite orbital
angular momentum of mechanical rotation can be transported into a
finite spin, which is known as the Barnett effect.  The Barnett effect
is an inverse phenomenon of the Einstein--de~Haas effect, and these
effects are manifestations of more general gyromagnetic effects as
summarized in Ref.~\cite{RevModPhys.7.129}.  Then, conversely, it
would be a natural anticipation to associate the CVE with a
relativistic extension of the Barnett effect.  However, there are
still some unresolved issues, and so this section must remain
speculative.

\subsubsection{Chiral Barnett effect}

The Barnett effect is quantified by a formula connecting the
magnetization, $\bM$, and the angular velocity vector, $\bomega$.  A
quick derivation of the formula is found in
Ref.~\cite{RevModPhys.7.129}, which can be readily understood in terms
of the energy shift.  The angular momentum, $\bJ$, makes an energy
shift by $\bJ\cdot\bomega$, which is equated to the magnetic energy
shift by $\bmu\cdot\bH_{\rm eff}$ with the magnetic moment $\bmu$ and
the effective magnetic field $\bH_{\rm eff}$.  The magnetization is
$\bM=\chi_B\bH_{\rm eff}$ with the magnetic susceptibility and the
magnetic moment is $\bmu=\gamma\bJ$ with the gyromagnetic ratio
$\gamma$.  Then, $\bH_{\rm eff}$ and $\bmu$ can be eliminated, that
is, the standard formula for the non-relativistic Barnett effect obtains as
\[
\framebox{
  \begin{minipage}{0.8\textwidth}
    Barnett Effect: A finite magnetization, $\bM$, along the rotation
    axis appears for a net charge neutral object spun with the angular
    velocity $\bomega$, which is formulated as
    \begin{equation}
      \bM = \frac{\chi_B}{\gamma}\bomega
    \end{equation}
  \end{minipage}
}
\]
The gyromagnetic ratio of free fermion reads as $\gamma=g\cdot q/(2m)$
with the electric charge $q$ and the mass $m$ together with a
$g$-factor.  Then, the magnetization is proportional to the mass, and
thus the Barnett effect becomes more prominent with larger mass, which
motivates the observation of the nuclear Barnett effect.

In contrast to heavy mass enhancement, in relativistic generalization
for light particles, the most subtle part is the decomposition of the
angular momentum into the orbital and the spin components.  What is
conserved as a N\"{o}ther current is the total angular momentum only,
which is immediately found from the Dirac Lagrangian as
\begin{equation}
  J^k = \epsilon^{ijk}\,\rmi\bar{\psi}\Bigl( \gamma^0 x^i\partial^j
  -\gamma^0 x^j\partial^i + \frac{1}{4}\gamma^0[\gamma^i , \gamma^j]
  \Bigr)\psi\,.
\end{equation}
It is quite natural, in a sense that the non-relativistic limit is
smoothly taken, to consider the first two terms involving the
coordinates as the orbital angular momentum and the rest as the spin.
Then, this identification leads to $S^k=\frac{1}{2}j_5^k$, that is, the axial
vector current itself is the spin.  With this hypothetical
identification, the spin expectation value is expressed for a given
distribution function $f(p)$ as studied in Ref.~\cite{Fukushima:2018osn}.  The
answer up to the first order in the $\omega$ expansion and also the
$\hbar$ expansion has two contributions:
\begin{equation}
  \langle\bS\rangle_{R/L} \simeq
  - \hbar^2 \frac{\bomega}{4}
  \int\frac{\rmd^3 p}{(2\pi)^3}\, f_{R/L}'(p)
  \mp \hbar \frac{\bomega\times\bx}{6}
  \int\frac{\rmd^3 p}{(2\pi)^3}\, p f_{R/L}'(p)\,.
\end{equation}
This expression uses a little sloppy notation.  The subscript $R/L$
denotes a contribution from one-handed fermion sector.  The first term
is a part of the CVE current as given in Eq.~\eqref{eq:cve_f}.  This
induced spin expectation value yields a finite magnetic moment, so
that the longitudinal magnetization is concluded, which is a natural
relativistic extension of the conventional Barnett effect.

\begin{figure}
  \centering
  \includegraphics[width=0.3\textwidth]{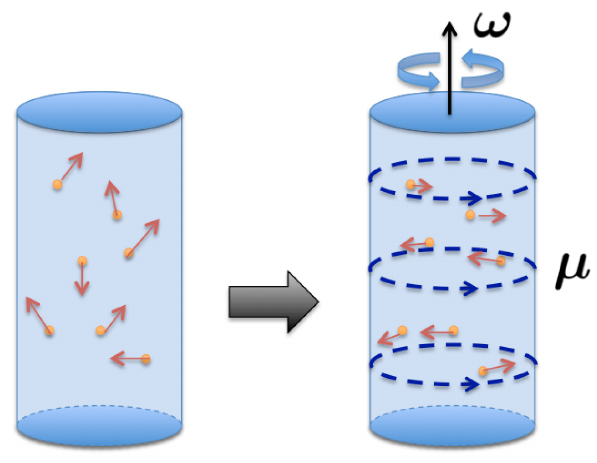}
  \caption{Sketch for the chiral Barnett effect: the spin of chiral
    fermions is aligned along the momentum direction.  For rotating
    system where fermions collectively have a net momentum on average,
    the spin expectation value has a transverse component.  Figure
    taken from Ref.~\cite{Fukushima:2018osn}.}
  \label{fig:barnett}
\end{figure}

The physical interpretation of the second term proportional to
$\bomega\times\bx$, which is denoted here by
$\langle\bS\rangle_\perp$, is highly nontrivial.  To clarify the
intuitive meaning, it would be useful to reexpress it as
\begin{equation}
  \langle\bS\rangle_\perp = \sum_{\pm} \mp \hbar\frac{\bomega\times\bx}{6}
  \int\frac{\rmd^3 p}{(2\pi)^3}\, p f'_{R/L}(p)
  = \frac{\hbar}{2}(\bomega\times\bx) n_5\,,
\end{equation}
where we see that $\bomega\times\bx$ represents the
transverse velocity of rotating matter.  Because right-handed
(left-handed) fermions have the spin alignment in parallel
(anti-parallel) to the momentum direction, therefore, this transverse
Barnett effect is an inevitable consequence from a combination of
chiral imbalance and rotation as graphically shown in
Fig.~\ref{fig:barnett}.  It should be noted that analogous nontrivial
effects as results of spin-rotation coupling have been predicted in a
different context~\cite{PhysRevLett.106.076601}.  So far, this
transverse Barnett effect for relativistic fermions is only a theory
conjecture, and more studies await.

\subsubsection{Chiral Einstein--de~Haas effect}

The essence for the Einstein--de~Haas effect is the angular momentum
conservation.  That is, the spin relaxation results in a finite amount
of the orbital angular momentum so that the total angular momentum
must be conserved.  Because the orbital angular momentum is associated
with mechanical rotation, such a transported orbital momentum causes a
physical rotation.  In this sense, the conservation law is robust, and
the relativistic extension should be straightforward.

To concretize the physical setup, let us suppose chiral matter at rest
(i.e., no rotation yet).  Then, a finite magnetic field is
adiabatically introduced into the system.  A finite spin expectation
value grows up in response to imposed magnetic field.  Actually, the
magnetic field itself applied to charged particles carries a finite
angular momentum as argued in Ref.~\cite{PhysRevLett.113.240404},
which makes the realization of conservation law a little complicated.
Once the spin operator in relativistic theory is identified as the
axial vector current, it is a classically conserved quantity for
chiral matter with $m=0$.  This means that, without $\calP$- and
$\calC\calP$-violating electromagnetic background, therefore, the
integrated spin averaged over the whole system should be vanishing,
and so the angular momentum of the magnetic field cannot be received
by the spin but may be maximally
transferred to the orbital angular momentum only, leading to an enhancement
in the Einstein--de~Haas effect.  This conjecture still awaits to be
formulated and quantified by field-theoretical investigations.  One
may think that the field-theoretical calculations are just
straightforward, but in practice, there appear nontrivial subtleties
in renormalization.  Up to now, to the best of my knowledge, no
successful calculation has been reported yet.

\section{Summary}
\label{sec:summary}

As excused in the beginning, this review is not a summarizing report
to cover all related works with balanced weights, but it is more like
a story along a history of my own research interest.  The order of
sections, subsections, subsubsections, therefore, has inevitability
for me as if the order of all historical events had definite
inevitability.  I must apologize for not mentioning all relevant
works, which is in any case impossible.

The magnetic catalysis is the most important phenomenon of chiral 
matter put in magnetic field.  Because there are already very nice
review articles to explicate the basics and the applications, I tried
to give only minimal but reproducible calculations.  The essential
point is that the vacuum in quantum field theories is more like a
medium in condensed matter physics and it is affected by environments
such as external magnetic fields.  There may be a phase transition
triggered by the magnetic field, and/or, the existing phase
transitions points are shifted by imposed magnetic field.  All these
have profound implications in nuclear physics as well as in other
fields.  To make this review pedagogical and self-contained, I could
not refer to some interesting ideas such as superconductivity of the
vacuum induced by the magnetic field~\cite{Chernodub:2011mc}, which is
an extremely interesting idea.  As mentioned above, usually, the
magnetic field influences medium properties, but conversely, the
magnetic field in Ref.~\cite{Chernodub:2011mc} plays a role of
medium.  In fact, such a view of the magnetic field as a medium can be
traced back to old discussions on the birefringence in the magnetic
field and/or the photon splitting problem under strong magnetic
field.  In this review I tried to spell out calculation procedures
especially in the early sections, so that readers can reproduce the
main results, and if necessary, can make use of the method for further
applications.  Because all algebraic elements are given, readers are
armed with theoretical tools, and ready to attack advanced problems as
listed above if interested.

I would like to stress one point.  For practical purposes Schwinger's
proper-time integral is the most convenient approach, but I preferred
to start with explicit solutions of the free Dirac equation.  This is
because the complete set of wave-functions contains more detailed
information than the propagator.  For example, to take account of
finite size effect, the most straightforward strategy is to solve the
Dirac equation under a boundary condition.  Also, for problems with
finite rotation, this rather brute-force method is useful to impose
the causality condition.  Such explicit construction of wave-functions
is also beneficial in order to think of new chiral effects.  In short,
anything interesting inherent to chirality is caused by helical
structures.  The matter chirality can indicate a helical motion, a
special optical setup may involve a helical magnetic profile, a screw
type rotation pumps charges up, etc.  Then, the wave-functions can
directly visualize the mechanism.

Everybody would agree that the inclusion of electric field is the next
step after treating magnetic fields.  The problem, however, becomes
much more difficult by orders of magnitude.  This is so because the
problem cannot be static any longer but should be dynamical as soon as
the external electric field is involved, which can be alternatively
understood as the sign problem in numerical Monte-Carlo simulations.
Thus, phase diagrams with an axis of external magnetic field can
exist, while matter exposed to external electric field cannot be
classified into equilibrated distinct phases.

A new element brought in by electric field is the particle production 
from quantum fluctuations.  In a sense the vacuum in quantum field 
theory should be not an empty state but more like a medium, as
emphasized above, and the particle production from the vacuum, known
as the Schwinger Mechanism, is a phenomenon of insulator-metal
transition in solid state physics.  Although inclusion of the magnetic
field on phase diagram research becase quite fashionable and a
countless number of publications appear especially for the strongly
interacting system of quarks and gluons, not many papers address the
electric effect so far.  This is because of technical difficulties;
the physics must be rich, but the calculation costs too much.  This is
the reason why I introduced, in this review, a relatively unknown
machinery for the electric calculus of Ref.~\cite{fradkin1991quantum}.
The inclusion of the electric field in real time deforms the
proper-time path, in a similar way as the finite-temperature field
theory.  In a sense we can say that the finite electric theory looks
like a holographic higher dimensional theory in which a quantum axis
is treated as if the temporal axis in the ordinary finite temperature
field theory.  This correspondence is quite suggestive;  it is a
longstanding puzzle whether the Schwinger Mechanism is thermal or
not.  Physics wise, the Schwinger Mechanism and the Unruh effect are
very similar, but the latter has a clear interpretation of the vacuum
as a thermal bath.  Why not the former?  Of course, a boring answer is
that the mass dimension of the electric field is not one but two.  It
should be worth mentioning that the proper time has a mass dimension
negative two, and this fact suggests that the thermal interpretation
may emerge in terms of the proper time, for which the formulation of
Ref.~\cite{fradkin1991quantum} might tell us a hint.

The last part of this review is about some ongoing research projects
on the relativistic generalization of classic themes, namely, the
Barnett effect and the Einstein--de~Haas effect.  Usually the Barnett
effect is proportional to the mass (since the gyromagnetic ratio
$\gamma$ is inversely proportional to the mass), heavier particles
like nuclei would be more desirable than lighter particles like
electrons for experimental observation.  However, at the same time,
the spin-orbit coupling is a relativistic effect, and it should be
enhanced for chiral fermions.  Then, it is a highly nontrivial
question whether such gyromagnetic effects are enhanced or suppressed
when the physical system transits from the non-relativistic to the
ultra-relativistic limit.  This part still awaits further theoretical
and experimental investigations.  In this way, this review is not a
standard review to rephrase established facts, but I dared to include
unmatured topics, hoping that they are seeds for rich harvest on the
next level in the near future.

\section*{Acknowledgments}
The author thanks
Hao-Lei~Chen,
Patrick~Copinger,
Shu~Ebihara,
Antonino~Flachi,
Francois~Gelis,
Yoshimasa~Hidaka,
Xu-Guang~Huang,
Dmitri~Kharzeev,
Tuomas~Lappi,
Kazuya~Mameda,
Takashi~Oka,
Jan~Pawlowski,
Shi~Pu,
Zebin~Qiu,
Harmen~Warringa
for discussions and collaborations.
The author learnt a lot from collaborations with them, and this review
is based on papers coauthored with them.

In the discussions on the magnetic catalysis I gave rather detailed
explanations in terms of the renormalization group.  The physical
contents are equivalent to direct calculations but such an intuitive
view to understand physics is quite interesting.  Actually, I learnt
this RG argument from Jan~Pawlowski who is an expert on the functional
RG approach, and then I felt like that my eyes were reopened to a new
streak of light.   I hope that some readers may find this RG argument
inspiring as I once did. 

In the presence of both electric and magnetic fields, interestingly, 
such background fields break $\calP$- and $\calC\calP$-symmetries, and
the particle--anti-particle symmetry is violated then.  This violation 
causes an unbalanced distribution of chirality, that is, a finite 
difference between right-handed and left-handed particles.  This kind 
of chiral unbalanced matter under magnetic field is an ideal device to 
realize the chiral magnetic effect.  Regarding the physical 
interpretation, since when Dmitri~Kharzeev, Harmen~Warringa, and I 
published a paper on the CME, it had been a puzzle how to distinguish 
the genuine current in real time and the Euclidean and thus static 
expectation value of the current operator.  For example, in lattice 
discretized numerical simulation as done by Arata~Yamamoto, it is
always possible to measure the current operator expectation value, but
such Euclidean formulation cannot be a proper description of real-time
phenomena.  After 10 years since our paper on the CME, I believe that,
thanks to collaboration with Patrick~Copinger and Shi~Pu, I eventually
reached a clear way to understand the difference between the
Minkowskian and the Euclidean quantities using the in- and out-states.
I would emphasize that using the chiral chemical potential, $\mu_5$,
which is a great invention by Harmen, is a very convenient bookkeeping
device, but, if controversies are fomented by $\mu_5$, an alternative
and more physical formulation with $\bE\cdot\bB$ would be helpful to
resolve confusions.

Analogy between the magnetic field and the rotation is another 
longstanding problem.  I learnt a lot about the Floquet theory from 
Takashi~Oka and was convinced that the Floquet engineering is such a
promising technique to test various ideas of relativistic quantum 
field theory including the quantum anomaly.  With this technique, even
theory fictions like axial vector potentials can be implemented in 
real experiments.  Now, the applications of synthetic magnetic field 
(aka, artificial gauge potential) have been a wide spreading subject, 
and it is not a radical idea at all to expect some Floquet engineering 
suited for studies on the CME and the analogous phenomena, e.g., the 
chiral vortical effect.  Along these directions I am (trying to)
keeping touch with Takashi and another Floquet expert,
Andr\'{e} Eckardt, from whom I learnt as much.

I could continue my personal memories endlessly, but now, let me
finally close the review, stating that this work was supported by JSPS
KAKENHI Grant No.\ 18H01211.


\bibliographystyle{utphys.bst}
\bibliography{mag}

\end{document}